\documentclass[12pt, a4paper]{article}
\usepackage{graphicx}
\usepackage{setspace,amsmath,amssymb,amsthm}
\usepackage{bm}
\usepackage{bbm}
\usepackage{enumitem,color}
\usepackage{mathrsfs}
\usepackage{physics}
\usepackage{fancybox}
\usepackage{ascmac}
\usepackage{algorithm}
\usepackage{algorithmic}
\usepackage{url}
\usepackage{natbib}
\usepackage{comment}

\usepackage[top=25mm,bottom=27mm,left=25mm,right=25mm]{geometry}

\newtheorem{prop}{Proposition}[section]
\newtheorem{thm}{Theorem}[section]
\newcommand{\E}[2]{\mathbb{E}_{#1}[#2]}
\newcommand{\Var}[2]{\mathrm{Var}_{#1}\left[#2\right]}

\numberwithin{equation}{section}

\title{Bayesian mixture modeling using a mixture of finite mixtures with normalized inverse Gaussian weights}
\author{Fumiya Iwashige and Shintaro Hashimoto \\
Department of Mathematics, Hiroshima University}
\date{}

\begin{document}
\maketitle
\doublespacing
\begin{abstract} 
In Bayesian inference for mixture models with an unknown number of components, a finite mixture model is usually employed that assumes prior distributions for mixing weights and the number of components. This model is called a mixture of finite mixtures (MFM). As a prior distribution for the weights, a (symmetric) Dirichlet distribution is widely used for conjugacy and computational simplicity, while the selection of the concentration parameter influences posterior inference. As a robust alternative to Dirichlet weights, we present a method based on a MFM with normalized inverse Gaussian weights. The motivation is similar to the use of normalized inverse Gaussian processes instead of Dirichlet processes in nonparametric Bayesian statistics. Introducing latent variables, the posterior computation is carried out using blocked Gibbs sampling without using the reversible jump algorithm. We also consider extensions to dynamic MFMs after clarifying the relationship with telescoping sampling. The performance of the proposed method is illustrated through some numerical experiments and real data examples, including clustering, density estimation, and community detection. Supplementary materials for this article are available online.
    
\end{abstract}


\section{Introduction}
\label{sec:intro}

We consider a finite mixture model with an unknown number of components:
\begin{align}
    f(y\mid\theta_{1},\dots,\theta_{M},\pi) = \sum_{j=1}^{M}\pi_{j}f(y\mid\theta_{j}) \label{mixture_model},
\end{align}
where $M \in \{1,2,\dots\}$ is the number of components, $y \in \mathbb{R}^{d}$ is a $d$-dimensional observation vector, and $\pi=(\pi_{1},\dots,\pi_{M})^{\top}$ is the mixing weight vector, satisfying  $\sum_{j=1}^{M}\pi_{j} = 1$ and $\pi_{j} > 0$. For $j=1,\dots, M$,  $f(\cdot\mid\theta_j)$ is a probability density or probability mass function parameterized by the component-specific parameter $\theta_j$ and is also called the kernel of the mixture model. Each observation is assumed to arise from one of the $M$ components, and each component is weighted based on its frequency of occurrence. Mixture models are important statistical models used in model-based clustering and density estimation because they can model complex data-generating distributions in random phenomena. Finite mixture models have been applied in many fields, for instance, sociology \citep{handcock2007model}, economics \citep{frohwirth2008model}, and genetics \citep{mclachlan2002mixture}, among others.  In applications, the determination of $M$ is a very important issue. Although various methods have been proposed for the selection of $M$, such as model selection criteria and hypothesis testing, using these criteria makes it difficult to quantify the uncertainty of $M$, and also introduces the bias that model selection is carried out. A comprehensive review of (finite) mixture models is provided by \cite{fruhwirth2006finite, mclachlan2019finite}. In this paper, we focus on the case where $M$ is finite and make a clear distinction between the number of components $M$ and the number of clusters $k$. That is, $M = k + M_{na}$, where $k$ is the number of components for which data are actually assigned, and $M_{na}$ is the number of empty components \citep[see also][]{argiento2022infinity}. 

Although we also estimate $M$ in this paper, there are two important issues associated with its inference. First, for any finite dataset, $M$ is not a well-defined quantity and cannot be directly inferred from the data. In certain application domains of mixture models, such as species sampling in biology \citep[see, e.g.,][]{argiento2014density}, it is even expected that data from some components may not be observed at all. Thus, given the data, we can only make predictions about $M$, and the prior on $M$ and the prior on the weight distribution play a crucial role. Second, the posterior distribution of $M$ is highly sensitive to model misspecification. 
Most consistency results for $M$ are derived under the assumption of model correctness \citep{nobile1994bayesian, miller2023consistency}. When attempting to infer $M$, it is critically important to ensure that the data-generating process is appropriately captured.

In Bayesian analysis, a finite mixture model that assumes prior distributions for the mixing weights and the number of components is usually employed \citep[see, e.g.,][]{nobile1994bayesian, miller2018mixture}. Such a model is also called a mixture of finite mixtures (MFM), and the model is often used, as well as infinite mixture models represented by Dirichlet process mixture models. Although the reversible jump algorithm \citep{green1995reversible, richardson1997bayesian} has been used to obtain samples from the posterior distribution based on an MFM, it faces significant computational and implementation challenges. Methods based on marginal likelihoods have also been proposed \citep[see, e.g.,][]{nobile2007bayesian}, while, as with reversible jump, the computational aspects are an issue. Recently, the use of nonparametric Bayesian methods for an MFM to estimate $M$ has attracted much attention. In \cite{miller2018mixture}, the exchangeable partition distribution is derived by marginalizing out $M$, and the restaurant process is constructed to overcome computational difficulties. Note that the result of Miller and Harrison (2018) has some connection with the Pitman-Yor process prior (see, e.g., Section 4.1 in \cite{fruhwirth2021generalized}). In \cite{fruhwirth2021generalized}, a generalized MFM in which the parameters of the Dirichlet distribution depend on $M$ is proposed. They showed that this model class may be regarded as a Bayesian nonparametric mixture outside the class of Gibbs-type priors, and proposed a novel sampling algorithm called telescoping sampling. In \cite{argiento2022infinity}, they developed an MFM framework based on a nonparametric Bayesian prior,
the normalized independent finite point process. By introducing a gamma latent variable, their approach encompasses a broad class of distributions on the simplex 
and yields an efficient blocked Gibbs sampling derived from the posterior random measure.

Most studies on MFMs employ a (symmetric) Dirichlet distribution for the mixing weights because of its analytical convenience. 
\cite{argiento2022infinity} primarily consider not only the Dirichlet distribution but also a family of simplex distributions constructed from $\sigma$-stable laws. 
However, the $\sigma$-stable example is treated as a central case, since their paper emphasizes that a restaurant process based computational algorithm can be constructed whenever the Laplace transform of the weight distribution is available. 

In this paper, we propose a new MFM and a generalized MFM based on the normalized inverse Gaussian distribution, with particular emphasis on the case in which the shape parameter is divided by $M$. We compare the proposed MFMs with the MFMs on the Dirichlet prior, focusing on the distributional differences. Although the normalized inverse Gaussian process has been studied for Bayesian nonparametric inference in infinite mixture models \citep{lijoi2005hierarchical}, normalized inverse Gaussian weights have received little attention in the MFM literature. We constructed a standard MFM with normalized inverse Gaussian weights by building on the results of \cite{argiento2022infinity}. We then developed an extended telescoping sampler that extends the original telescoping sampler to general distributions on the simplex and used it to construct generalized MFMs with these mixing weights. To derive the extended telescoping sampler, we clarified the relationship between the telescoping sampler and the blocked Gibbs sampler. To the best of our knowledge, this is the first study to formally clarify the relationship between the results of \cite{fruhwirth2021generalized} and those of \cite{argiento2022infinity}. By developing the extended telescoping sampler based on this connection, we provide new insights into the efficiency of MFM computations. These contributions represent a novel advancement in expanding the MFM toolbox and bridging previously distinct methodologies.

This paper is organized as follows. In Section \ref{sec:2}, we provide a brief review of MFMs and their equivalent representation using discrete probability measures. We also discuss the data augmentation and the conditional posterior distribution of $M_{na}$, which are crucial for constructing the computational algorithm. In Section \ref{sec:3}, we present the proposed method as well as the posterior computation algorithm. In Section \ref{sec:4}, we illustrate some numerical studies to compare the proposed method with existing methods. R code implementing the proposed method is available at GitHub repository \url{https://github.com/Fumiya-Iwashige/MFM-Inv-Ga}

\section{Mixture of finite mixtures}
\label{sec:2}
We present a brief overview of an MFM and its equivalent representation of the discrete probability measure. The proposed model and the posterior computation algorithm presented in the next section are largely based on the basic model presented in this section. 

\subsection{Formulation}
\label{subsec:2.1}
Let each observation $Y_{1},\dots,Y_{n}$ be univariate or multivariate in a Euclidean space. An MFM is a statistical model defined by the following hierarchical representation:
\begin{align}\label{MFM}
\begin{split}
&M-1 \sim q_{M},\quad q_{M}\ \text{is a probability mass function (p.m.f.) on}\ \{0,1,2,\dots\},\\
&\pi\mid M \sim P_{\pi}(\pi\mid M),\\
&c_{i}\mid M,\pi \sim \mathrm{Categorical}_{M}(\pi), \quad i=1,\dots,n,\\
&\tau_{m}\mid M \overset{\mathrm{i.i.d}}{\sim} p_{0}(\tau),\quad m=1,\dots,M,\\
&\theta_{i}\mid c_{i},\tau_{1:M} \overset{\mathrm{ind}}{\sim} \delta_{\tau_{c_{i}}}(\theta_{i}),\quad i=1,\dots n,\\
&Y_{i}\mid \theta_{i} \overset{\mathrm{ind}}{\sim} f(y_{i}\mid \theta_{i}), \quad i=1,\dots, n,
\end{split}
\end{align}
where $f(y\mid\theta_i)$ is a parametric model with parameter $\theta_i$, $\delta_{x}(\cdot)$ is a point mass at $x$, and $\tau_m$ is an element of a vector $\tau_{1:M} = (\tau_{1},\dots,\tau_{M})^{\top}$. $p_{0}(\tau_{1:M})$ is a prior density function of the parameter in mixture components. The parameter space is denoted by $\Theta \subset \mathbb{R}^{d}$. Each $c_{i}$ is a latent allocation that indicates to which component each observation is allocated. Given the number of components $M$, $P_{\pi}(\pi\mid M)$ is a probability distribution on the $M$-dimensional unit simplex, in that this is the prior distribution for the mixing weights. 

Let $h$ be a probability density function (p.d.f.) on $\mathbb{R}_{+}$. We focus on the case where the mixing weights can be further hierarchically expressed as follows:
\begin{align}\label{simplex}
    \begin{split}
        &S_{m}\mid M \overset{\mathrm{ind}}{\sim} h,\\
        &\pi=\left(\frac{S_{1}}{T},\dots,\frac{S_{M}}{T}\right) \sim P_{\pi}(\pi\mid M),\quad T = \sum_{m=1}^{M}S_{m},
    \end{split}
\end{align}
where $S_1,\dots,S_M$ are unnormalized weights and $\pi$ is a (normalized) mixing weight vector.  
This representation is the most basic way to generate a probability distribution on the $d$-dimensional unit simplex $\mathbb{S}_{d} := \{w=(w_{1},\dots,w_{d});w_{i} \geq 0, \sum_{i=1}^{d}w_{i} = 1\}$. One of the most famous examples of a distribution on $\mathbb{S}_{d}$ is the Dirichlet distribution. If $S_i\sim \text{Gamma}(\gamma_{i}, 1)$ in \eqref{simplex}, $\pi\sim \text{Dirichlet}(\gamma_{1},\dots,\gamma_{d})$, where $\text{Gamma}(a, b)$ is the gamma distribution with shape parameter $a>0$ and rate parameter $b>0$, and $\text{Dirichlet}(\gamma_{1},\dots,\gamma_{d})$ is the Dirichlet distribution with parameter $(\gamma_1,\dots,\gamma_d)$. 
When we set $\gamma_1=\cdots=\gamma_d \ (=\gamma > 0)$, we have the symmetric Dirichlet distribution $\text{Dirichlet}(\gamma,\dots,\gamma)$, and this distribution is often used as a prior distribution for the mixing weight vector in the MFM framework, because of conjugacy with categorical distributions and simplicity of computation. The symmetric structure of $\text{Dirichlet}(\gamma,\dots,\gamma)$ is also essential for marginalizing out $M$ and deriving the exchangeable partition probability function (EPPF). 

Also, if the hyperparameters of $h$ depend on $M$ in \eqref{simplex}, the model is referred to as a generalized MFM:
\begin{align}\label{eq:GMFM}
    \begin{split}
        &S_{m}\mid M \overset{\mathrm{ind}}{\sim} h_M,\quad m=1,\dots,M,\\
        &\pi=\left(\frac{S_{1}}{T},\dots,\frac{S_{M}}{T}\right) \sim P_\pi(\pi\mid M),\quad T = \sum_{m=1}^{M}S_{m}.
    \end{split}
\end{align}
In particular, if the shape parameters are divided by $M$, the model is referred to as a \emph{dynamic} MFM (DMFM). The Dirichlet case, where $h_M$ is $\text{Gamma}(\gamma/M, 1)$, is discussed in \cite{fruhwirth2021generalized}. On the other hand, \cite{fruhwirth2021generalized} referred to the MFM defined by \eqref{MFM}, in which the distributional parameters of $S_m$ do not depend on $M$, as a \emph{static} MFM. In this paper, we simply refer to such a static MFM as an MFM.

However, it is known that the estimation result is sensitive to the choice of the shape parameter $\gamma$. For example, \cite{miller2018mixture} recommended using $\gamma=1$ as a default choice and the value works well in many cases. On the other hand, in a sparse MFM, dividing by a large $M$ effectively corresponds to using a small value of $\gamma$. Moreover, because $\gamma$ strongly influences both the prior distribution of the number of clusters and the induced partition distribution, it is necessary to assign an appropriate prior to $\gamma$. For further discussion of broader prior specifications, see \cite{greve2022spying}.

\subsection{Equivalent representations using discrete probability measures}
\label{subsec:2.2}
Following \cite{argiento2022infinity}, we give an equivalent representation of the MFM. We can construct a discrete measure $P(\cdot) = \sum_{m=1}^{M}\pi_{m}\delta_{\tau_{m}}(
\cdot)$ in the parameter space $\Theta$, almost surely, where $M$, $\tau$, and $\pi$ are realizations from the distributions $q_{M}$, $p_{0}$, and $P_{\pi}$, respectively. Let $\theta_{1},\dots,\theta_{n}$ be random variables according to $P$. Then, the model \eqref{MFM} is equivalent to the following hierarchical representation using a random measure $P$:
\begin{align}\label{MFM_non}
    \begin{split}
        &Y_{i}\mid\theta_{i} \overset{\mathrm{ind}}{\sim} f(y_{i}\mid\theta_{i}), \quad i=1,\dots, n,\\
        &\theta_{1},\dots,\theta_{n}\mid P \overset{\mathrm{i.i.d}}{\sim} P,\\
        &P \sim \mathcal{P}(q_{M}, h, p_{0}),
    \end{split}
\end{align}
where $\mathcal{P}$ is a probability distribution of $P$ with parameters $q_{M}$, $h$, and $p_{0}$. The representation \eqref{MFM_non} is the MFM described in the same framework as infinite mixtures. If we replace $\mathcal{P}$ with the Dirichlet process, \eqref{MFM_non} represents the well-known Dirichlet process mixture models \citep[e.g.,][]{escobar1995bayesian}. If we replace $\mathcal{P}$ with the normalized inverse Gaussian process, \eqref{MFM_non} represents the normalized inverse Gaussian process mixture models \citep{lijoi2005hierarchical}. For MFMs, \cite{argiento2022infinity} proposed a normalized independent finite point process (Norm-IFPP), which is a class of flexible prior distributions for $P$. We employ the representation \eqref{MFM_non} with a Norm-IFPP. The advantages are as follows. We can directly estimate $M$ and $k$. Furthermore, an efficient Gibbs sampler can be constructed by incorporating the data augmentation with a latent Gamma random variable. This data augmentation enables us to overcome the lack of conjugacy in the categorical distribution. Instead of using the probability density function of $P_{\pi}$, we can use the density function of $h$. This is the key to building an efficient MCMC algorithm for the proposed model. The details of a Norm-IFPP and a general independent finite point process (IFPP) are given in \cite{argiento2022infinity}.

\subsection{Data augmentation and conditional distribution of $M_{na}$}\label{sec:2.3}
We here illustrate the data augmentation and conditional posterior distribution of $M_{na}$. We introduce the data augmentation in models (\ref{MFM}) and (\ref{simplex}). This technique is employed in \cite{james2009posterior} and \cite{argiento2022infinity}. Let $\mathcal{M}_{a}$ and $\mathcal{M}_{na}$ be the allocated and unallocated index sets, respectively. The conditional joint distribution of $\tau = (\tau_{1},\dots,\tau_{M})^\top$ and $S = (S_{1},\dots,S_{M})^\top$ given $M$ and a label vector $c$ is
\begin{align}
    p(S, \tau\mid M, c) 
    &\propto \left(\prod_{i=1}^{n}\frac{S_{c_{i}}}{T}\right) \left(\prod_{m=1}^{M}h(S_{m})p(\tau_{m})\right) \notag\\
    &=\left(\prod_{m \in \mathcal{M}_{a}}\left(\frac{S_{m}}{T}\right)^{n_{m}}h(S_{m})p(\tau_{m})\right)\left(\prod_{m \in \mathcal{M}_{na}}h(S_{m})p(\tau_{m})\right), \notag
\end{align}
where $n_{m}=\#\{i:c_{i} = m\}$ and $T=\sum_{m=1}^M S_m$, which depends on $S$. Furthermore, using $U_n \mid T \sim \mathrm{Gamma}(n, T)$, we obtain
\begin{align*}
    p(S,\tau,U_n\mid M,c) \propto u^{n-1}\left(\prod_{m \in \mathcal{M}_a}e^{-S_{m}U_n}S_{m}^{n_{m}}h(S_{m})p(\tau_{m}) \right)\left(\prod_{m \in \mathcal{M}_{na}}e^{-S_{m}U_n}h(S_{m})p(\tau_{m})\right).
\end{align*}
Thus, the conditional distribution of each $S_{m}$ is 
\begin{align}
    &S_{m}\mid U_{n}=u,M=m,c \overset{\mathrm{ind}}{\sim} e^{-S_{m}u}S_{m}^{n_{m}}h(S_{m}),\quad m \in \mathcal{M}_{a}, \label{full_s_m_a}\\
    &S_{m}\mid U_{n}=u,M=m,c \overset{\mathrm{i.i.d}}{\sim} e^{-S_{m}u}h(S_{m}),\quad m \in \mathcal{M}_{na}. \label{full_s_m_na}
\end{align}
If it is easy to generate random variables from the distributions in (\ref{full_s_m_a}) and (\ref{full_s_m_na}), an efficient Gibbs sampling algorithm can be constructed. For example, when the prior distribution of the unnormalized weight is an inverse Gaussian distribution, (\ref{full_s_m_a}) and (\ref{full_s_m_na}) are generalized inverse Gaussian distributions. Introducing $U_{n}$, the update of the variable $\pi$ is replaced by the update of the variable $S_{m}$. Thus, the selection of $h$ is essential in MCMC updates.

Under this data augmentation, the conditional distribution of $M_{na}$ is established in Theorem 5.1 in \cite{argiento2022infinity}. This theorem states that if $\mathcal{P}(q_{M}, h, p_{0})$ follows a Norm-IFPP, then the posterior distribution of the random measure $P$, given $U_{n}$, is a superposition of a finite point process with fixed points and an IFPP. The IFPP characterizes the process of unallocated jumps, where the discrete probability distribution that serves as its parameter corresponds to the distribution of $M_{na}$. The conditional distribution of $M_{na}$ is given by
\begin{align}
    \mathbb{P}(M_{na} = m\mid\theta,U_{n}=u,k) \propto \frac{(m+k)!}{m!}\psi(u)^{m}q_{M}(m+k), \quad m=1,2,\dots, \label{cond_M_na}
\end{align}
where $\theta = (\theta_{1},\dots,\theta_{n})^\top$, $k$ is the number of clusters (unique values of $\theta$), and $\psi$ is the Laplace transform of $h$. We sample $M_{na}$ from (\ref{cond_M_na}) in the MCMC algorithm. Then, we straightforwardly obtain $M$ by adding $k$ to $M_{na}$.

\section{Methodology}
\label{sec:3}
In this section, we propose a mixture of finite mixtures with normalized inverse Gaussian weights. The posterior inference based on the proposed model can be efficiently implemented using the blocked Gibbs sampling proposed by \cite{argiento2022infinity}. 

\subsection{Mixture of finite mixtures with normalized  inverse Gaussian weights}
\label{subsec:3.1}
We propose a mixture of finite mixtures with normalized inverse Gaussian weights (denoted by MFM-IGau), where the notation explicitly specifies $h$ because it is essential for computation. Our proposal model only requires (2.2) to be
\begin{align*}
    &S_{m}\mid M \overset{\mathrm{ind}}{\sim} \text{IGau}(\alpha, 1),\quad \alpha > 0,\\
    &\pi = \left(\frac{S_{1}}{T},\dots,\frac{S_{M}}{T}\right) \sim \text{NIGau}(\alpha,\dots,\alpha),
\end{align*}
where $\text{IGau}(\alpha, \beta)$ is the inverse Gaussian distribution with shape parameter $\alpha > 0$ and scale parameter $\beta > 0$, and $\text{NIGau}(\alpha_{1},\dots,\alpha_{d})$ is the normalized inverse Gaussian distribution with parameters $\alpha_{i} > 0$ for $i=1,\dots,d$. The probability density function of a random variable $S$ according to $\text{IGau}(\alpha, \beta)$ is given by
\begin{align}\label{Inv-Ga_pdf}
      h(s) = \frac{\alpha}{\sqrt{2\pi}}s^{-3/2}\exp\left(-\frac{1}{2}\left(\frac{\alpha^2}{s} + \beta^2s\right) + \beta\alpha\right), \quad s>0, 
\end{align}
where $\alpha>0$ is the shape parameter and $\beta>0$ is the scale parameter. 
The mean and variance of $S$ are given by
\begin{align}\label{Inv-Ga_pdf_ex_var}
    \E{}{S} = \frac{\alpha}{\beta},\quad \Var{}{S}=\frac{\alpha}{\beta^3}.
\end{align}
Let $\psi_{\text{IGau}}(u)$ be the Laplace transforms  of $\text{IGau}(\alpha, 1)$. $\psi_{\text{IGau}}(u)$ is the following:
\begin{align}
    \psi_{\text{IGau}}(u) = \exp\left(\alpha\left(1 - \sqrt{1 + 2u}\right)\right),\quad u \geq 0\label{Inv-Ga_psi}.
\end{align}
Also, cumulant $\kappa_{\mathrm{IGau}}(n,u) := (-1)^{n}\frac{d^{n}}{du^{n}}\psi_{\mathrm{IGau}}(u)$ is 
\begin{align}
    &\kappa_{\mathrm{IGau}}(u;n) =\alpha^{n}\psi_{\text{IGau}}(u)(1+2u)^{-n/2}\frac{K_{n-1/2}\left(\alpha\sqrt{1+2u}\right)}{K_{1/2}\left(\alpha\sqrt{1+2u}\right)},\label{eq:cum_IGau}
\end{align}
where $K_m$ is the modified Bessel functions  of the second kind of order $m$. The derivation of \eqref{eq:cum_IGau} is provided in the Supplementary Material. From \cite{lijoi2005hierarchical}, the probability density function of the normalized inverse Gaussian distribution with parameters $\alpha_{i} > 0,\ i=1,\dots,d$ is given by 
\begin{align}\label{Norm_Inv_Ga_pdf}
        \begin{split}
            f(\pi) = &\frac{e^{\sum_{i=1}^{d}\alpha_{i}}\prod_{i=1}^{d}\alpha_{i}}{2^{d/2-1}\pi^{d/2}} \times K_{-d/2}\left(\sqrt{\mathcal{A}_{d}(\pi_{1},\dots,\pi_{d-1})}\right)\\
        &\times \left(\pi_{1}^{3/2}\cdots \pi_{d-1}^{3/2}\left(1 - \sum_{i=1}^{d-1}\pi_{i}\right)^{3/2} \times \mathcal{A}_{d}(\pi_{1},\dots,\pi_{d-1})^{d/4}\right)^{-1},
        \end{split}
\end{align}
where $\mathcal{A}_{d}(\pi_{1},\dots,\pi_{d-1}) = \sum_{i=1}^{d-1}\alpha_{i}^2/\pi_{i} + \alpha_{d}^{2}\left(1-\sum_{i=1}^{d-1}\pi_{i}\right)^{-1}$. 

The normalized inverse Gaussian distribution is less informative and more robust to parameter specification than the Dirichlet distribution. In the context of a DMFM, shape parameters such as $\gamma/M$ or $\alpha/M$ become small, and information measures such as entropy and the Gini index of $\mathrm{Dirichlet}(\gamma/M, \dots, \gamma/M)$ also tend to be very small. Therefore, it is desirable to incorporate learning for $\gamma$. In contrast, those of $\mathrm{NIGau}(\alpha/M, \dots, \alpha/M)$ do not decrease as easily, so stable posterior inference can be achieved regardless of the choice of $\alpha$. This fact is confirmed through the numerical experiment in Section~4.1.1. Further analysis of the Gini index is provided in the Supplemental Material. Also, in the context of infinite mixtures, it is known that the prior on the number of clusters induced by the normalized inverse Gaussian process has heavier tails than that induced by the Dirichlet process \citep{lijoi2005hierarchical}. It is expected that this robustness carries over to the finite mixture setting as well. These provide the main motivations for using the normalized inverse Gaussian distribution as mixing weights.

Due to the breakdown of conjugacy with the categorical distribution, constructing an efficient Gibbs sampler is challenging. Because of the complexity of (\ref{Norm_Inv_Ga_pdf}), the calculation in \cite{miller2018mixture} is intractable: constructing an MCMC algorithm based on the restaurant process by marginalizing out $M$ is extremely challenging. To overcome this difficulty, we use the data augmentation technique and the sampling method proposed by \cite{argiento2022infinity}. As a result, it is no longer necessary to deal directly with the normalized inverse Gaussian distribution \eqref{Norm_Inv_Ga_pdf}, and the simple MCMC algorithm can be constructed using only the inverse Gaussian distribution \eqref{Inv-Ga_pdf} and its Laplace transform.

\subsection{Posterior computation}
\label{subsec:3.2}
In this section, we present a fast and efficient posterior computation algorithm for the proposed method. To this end, we adopt the blocked Gibbs sampling scheme in \cite{argiento2022infinity}. Moreover, we discuss the blocked Gibbs sampling in detail and explain the computational advantages of using inverse Gaussian weights.

Let $\eta$ and $\rho$ be hyperparameters of $q_{M}$ and $h$, respectively, and $p(\cdot)$ be a joint prior density function except for $h$ and $q_{M}$. $n_{m}=\#\{i;c_{i}=m\}$ is the size of the $m$th cluster. $\psi(u)$ is the Laplace transform of $h$, which is defined by $\psi(u) := \E{}{e^{-uS_{m}}}$ for $u \geq 0$. $\kappa(u;n)$ is the cumulant of $h$, which is defined by $\kappa(u;n) := (-1)^{n}\frac{d^{n}}{du^{n}}\psi(u)$. Note that $\psi(u)$ and $\kappa(u;n)$ depend on $\rho$.

\begin{algorithm}[t]
\caption{Blocked Gibbs sampling \citep{argiento2022infinity}}
\label{alg}
\begin{algorithmic}[htbp]
\STATE\textbf{Step 0.}\ Set initial values.\\
\STATE\textbf{Step 1.}\ Sample $U_{n}$ from $\mathrm{Gamma}(n, T)$,\ where $T = \sum_{m=1}^{M}S_{m}$.
\STATE\textbf{Step 2.}\ Sample $c_{i}$, for $i=1,\dots,n$, from the following discrete probability distribution defined by each probability:
\[
\mathbb{P}(c_{i} = j\mid\mathrm{rest}) \propto S_{j}f(y_{i}\mid\tau_{j}),\quad j=1,\dots,M.
\]
\STATE\textbf{Step 3.}\ Sample the hyperparameter $\eta$ of the $q_{M}$ from 
\[
p(\eta\mid\mathrm{rest}) \propto \Psi(u, k)p(\eta),\quad \Psi(u,k) := \sum_{m=0}^{\infty}\frac{(m+k)!}{m!}\psi(u)^{m}q_{M}(m+k).
\]
\STATE\textbf{Step 3'.}\ Sample the hyperparameter $\rho$ of the $h$ from 
\[
p(\rho\mid\mathrm{rest}) \propto \Psi(u, k)\prod_{j=1}^k\kappa(u;n_j)p(\rho).
\]
\STATE\textbf{Step 4.}\ Sample $M_{na}$ from the following discrete distribution:
\[
\mathbb{P}(M_{na}=m\mid\mathrm{rest}) \propto \frac{(m + k)!}{m!}\psi(u)^{m}q_{M}(m + k),\quad m=0,1,\dots.
\]
\STATE\textbf{Step 5.}\ Sample $S_{m}$, for $m=1,\dots,k$, from $e^{-S_{m}u}S_{m}^{n_{m}}h(S_{m}).$
\STATE\textbf{Step 6.}\ Sample $\tau_{m}$, for $m=1,\dots,k$, from
\[
p(\tau_{m}\mid\mathrm{rest}) \propto \left\{\prod_{i;\theta_{i}=\tau_{m}}f(y_{i}\mid\tau_{m})\right\}p(\tau).
\] 
\STATE\textbf{Step 7.}\ Sample $S_{m_{na}}$, for $m_{na}=k+1,\dots,M_{na}+k$, from $e^{-S_{m}u}h(S_{m}).$
\STATE\textbf{Step 8.}\ Sample $\tau_{m_{na}}$, for $m_{na}=k+1,\dots,M_{na}+k$, from the prior $p(\tau)$.
\end{algorithmic}
\end{algorithm}

We summarize the algorithm in Algorithm \ref{alg}. 
The point of this algorithm is the data augmentation through the latent variable such as $U_{n}\mid T \sim \mathrm{Gamma}(n, T)$, where $n$ is the sample size and $T=\sum_{m=1}^{M}S_{m}$. It is important to sample the number of empty components $M_{na}$ from (\ref{cond_M_na}) in Step 4. This step allows for direct sampling of $M$ by adding $k$ and label variables to be updated as in the finite mixture model with given $M$ in step 2. The update is the same as the telescoping sampling proposed by \cite{fruhwirth2021generalized}, and the method is more efficient than the classical restaurant process. However, in implementation, it is important to determine $q_M$ so that the series $\Psi(u,k)$ in Steps 3 and 4 can be written analytically and random variables can be easily generated from the full conditional distributions of $\eta$ and $M_{na}$. In Steps 5 and 7, the allocated and unallocated weights are updated based on \eqref{full_s_m_a} and \eqref{full_s_m_na}, respectively. The prior $h$ of $S_m$ is discussed in detail in the next paragraph. In Steps~7 and~8, the assigned unnormalized weights and kernel parameters are updated. When $M_{na}$ sampled in Step 4 is greater than or equal to $1$, Steps 7 and 8 are executed.

To implement the blocked Gibbs sampling, at least two conditions must be satisfied: 1) both the probability density function and the Laplace transform must be available in closed form, and 2) each $S_m$ can be easily generated from \eqref{full_s_m_a} and \eqref{full_s_m_na}. Additionally, it is desirable that the cumulants of $h$ can be expressed in closed form when updating $\eta$ or when extending the (static) MFM to the DMFM. When $S_m$ follows the $\sigma$-stable distribution, the blocked Gibbs sampling, in general, cannot be used, as the probability density function is available in integral form. Alternatively, a marginal Gibbs sampler based on the restaurant process is required, as employed in \cite{argiento2022infinity}.

The inverse Gaussian distribution satisfies all of the conditions. The probability distribution, the Laplace transform, and the cumulant are given by \eqref{Inv-Ga_pdf}, \eqref{Inv-Ga_psi}, and \eqref{eq:cum_IGau}, respectively. If $h$ is the $\text{IGau}(\alpha, 1)$, the full conditional distribution of $S_m$ is the generalized inverse Gaussian distribution. From (\ref{full_s_m_a}) and (\ref{full_s_m_na}), we have
\begin{align*}
    &S_{m}\mid U_{n}=u,M=m,c \overset{\mathrm{ind}}{\sim} \mathrm{GIG}(2u+1, \alpha^2, n_{m}-1/2),\quad m \in \mathcal{M}_{a} ,\\
    &S_{m}\mid U_{n}=u,M=m,c \overset{\mathrm{i.i.d}}{\sim} \mathrm{GIG}(2u+1, \alpha^2, -1/2),\quad m \in \mathcal{M}_{na},
\end{align*}
where $\mathrm{GIG}(a,b,c)$ is the generalized inverse Gaussian distribution with $a > 0, b > 0, c \in \mathbb{R}$.

\subsection{Prior distributions for the number of mixture components}
\label{subsec:3.3}
Assume that $M-1$ follows a discrete probability distribution with the support $\{0,1,2,\dots\}$. In this paper, we consider $\text{Poisson}(\Lambda)$ ($\Lambda > 0$) for $M-1$, because the constraints in Steps 3 and 4 are satisfied. In fact, from \cite{argiento2022infinity}, we have
\begin{align}\label{Psi_poi}
    \Psi(u,k) = \Lambda^{k-1}(\Lambda\psi(u) + k)\exp(\Lambda(\psi(u)-1)).
\end{align}
Assuming $\Lambda \sim \text{Gamma}(a_{\Lambda}, b_{\Lambda})$ for  $a_{\Lambda}, b_{\Lambda} > 0$, the full conditional distributions of $\Lambda$ and $M_{na}$ are given by 
\begin{align*}
    p(\Lambda\mid\mathrm{rest}) \propto \psi(u)(k+a_{\Lambda}-1)&\mathrm{Gamma}(k+a_{\Lambda}+1, 1-\psi(u)+b_{\Lambda}) \\
    &+ k(b_{\Lambda}+1-\psi(u))\mathrm{Gamma}(k+a_{\Lambda}, 1-\psi(u)+b_{\Lambda}),\\
    \mathbb{P}(M_{na}=m\mid\mathrm{rest}) \propto \Lambda\psi(u)&\textrm{Shifted-Poisson}(\Lambda\psi(u), 1) + k\mathrm{Poisson}(\Lambda\psi(u)),
\end{align*}
respectively, where $\text{Shifted-Poisson}(\Lambda, t)$ is $\text{Poisson}(\Lambda)$ shifted by $t$. Furthermore, if $b_\Lambda$ is hierarchically assigned a beta-prime prior $\text{Betaprime}(a_p, b_p), a_p, b_p > 0$, then the marginal prior of $M-1$ follows a beta-negative binomial distribution $\text{BNB}(a_\Lambda, a_p, b_p)$. 
\cite{fruhwirth2021generalized} recommend using $\text{BNB}(1, 4, 3)$ under Dirichlet weights. It should be noted, however, that updating $b_\Lambda$ requires the Metropolis-Hastings algorithm. Therefore, various discrete distributions can be incorporated into the prior for $M-1$ by using a Poisson-based hierarchical structure.

\subsection{Specification of kernels}
\label{sec:3.4}
The choice of kernel is important, and the appropriate kernel must be selected for the purpose. In this paper, although we do not discuss the details of the selection of kernels, we present some famous kernels for the sake of completeness. 

\subsubsection{Cluster analysis and density estimation}
\label{sec:3.4.1}
One of the most famous and useful kernels is the (multivariate) normal kernel $N(y\mid\mu,\sigma^2)$. In the univariate case, we often use the normal-inverse gamma model as a prior distribution of $\tau=(\mu,\sigma^2)$: $\mu \mid \sigma^2 \sim N(m_0, \sigma^2/\eta),\quad \sigma^2 \sim \text{IG}(c_0, C_0)$, where $\ m_{0} \in \mathbb{R}$, $\eta,\ c_{0},\ C_{0} > 0$. The parameter $\eta$ is called a smoothing parameter and plays an important role in density estimation. It is possible to include a hierarchical prior for $\eta$. In the multivariate case, the hierarchical normal inverse Wishart model is $\mu \sim N_r(b_0, B_0),\Sigma^{-1} \mid C \sim W(c_0, C), C \sim W(g_0, G_0)$, where $b_0 \in \mathbb{R}^r$, $B_0$ is an $r \times r$ covariance matrix, $c_0, g_0 > r - 1$, and $G_0$ is an $r \times r$ positive definite symmetric matrix. We used normal kernels in later numerical experiments in Sections \ref{sec:4.1} and \ref{sec:4.2} for clustering and density estimation. Of course, other kernels can also be used depending on the purpose \citep[see, e.g.,][]{fruhwirth2010bayesian}.

\subsubsection{Network analysis}
\label{subsec:3.4.2}
As an application of the proposed method, we perform community detection on network data. Community detection is the task of identifying dense subclasses in network data, and corresponds to clustering and estimating the number of components, called the number of communities in network analysis. Note that the number of components of finite mixture models is equivalent to the number of communities in the network, and both are denoted $M$. Estimating the number of communities is an important problem and various methods have been proposed \citep{shi2000normalized, white2005spectral}. From a model-based perspective, it is essentially the same as estimating the number of components in a finite mixture model, and we can apply an MFM. The stochastic block model is a famous statistical model of network data \citep{holland1983stochastic} which assumes a stochastic block structure behind it and specifies the community structure by estimating the probability of edges being drawn between each group. \cite{geng2019probabilistic} proposed a stochastic block model based on an MFM with Dirichlet weights, and also constructed a similar algorithm to \cite{miller2018mixture}.

An MFM can be easily applied to community detection by modifying the kernel. Data $y$ are replaced by the adjacency matrix $A \in \mathbb{R}^{n \times n}$, where $A = (A_{ij}) \in \{0, 1\}^{n \times n}$ and $n$ is the number of nodes. When $A_{ij} = 1$, this indicates that an edge is drawn from the $i$th node to the $j$th node, and when $A_{ij} = 0$, it is not. For simplicity, we assume that the adjacency matrix is undirected and does not have any self-loops, in that $A_{ij} = A_{ji}$ and $A_{ii} = 0$, where $1 \leq i < j \leq n$. The stochastic block model is formulated as follows:
\begin{align}\label{bi_SBM}
    \begin{split}
        &A_{ij}\mid Q,M \sim \text{Bernoulli}(\theta_{ij}),\quad \theta_{ij} =Q_{c_{i}c_{j}},\quad 1 \leq i < j \leq n,\\
        &Q_{rs} \sim \text{Beta}(a_{Q}, b_{Q}),\quad 1 \leq r \leq s \leq M,
    \end{split}
\end{align}
where $a_{Q},\ b_{Q} > 0$ and $M$ is the number of communities. $Q=(Q_{rs}) \in [0,1]^{M \times M}$ is a symmetric matrix and defines the stochastic block structure of a network. Each element $Q_{rs}$ represents the probability that an edge is drawn between any node belonging to the community labeled $r$ and any node belonging to the community labeled $s$. To perform community detection for the proposed model, we just set the Bernoulli likelihood as the kernel.

\subsection{Evaluation of the number of empty components}
\label{subsec:3.5}
\begin{figure}[htbp]
    \centering
    \includegraphics[width=\linewidth]{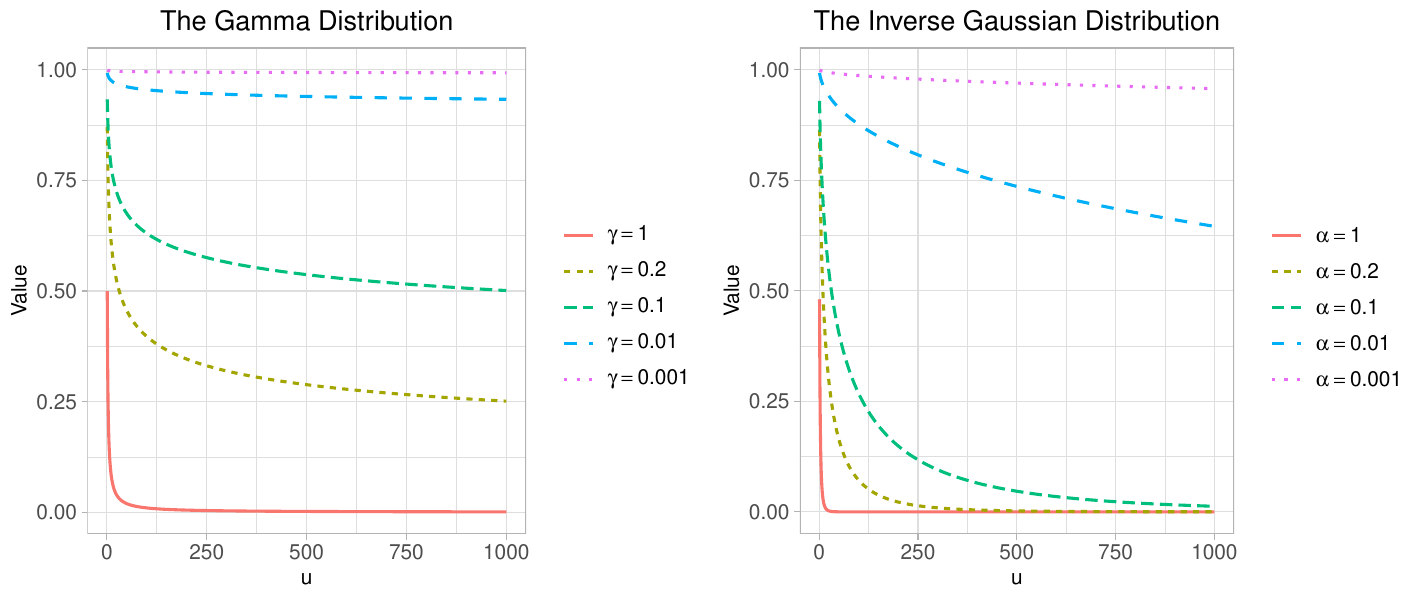}
    \caption{For shape parameters $\alpha=\gamma=1$ (solid), $0.2$ (dashed),
$10^{-1}$ (long-dashed), $10^{-2}$ (dot-dashed), and
$10^{-3}$ (dotted), the left panel shows the Laplace transforms of
$\mathrm{Gamma}(\gamma,1)$, while the right panel shows those of
$\mathrm{IGau}(\alpha,1)$.}
    \label{fig:laplace}
\end{figure}

 For the full conditional distribution of $M_{na}$, the following inequality holds, where $M-1\mid\Lambda \sim \mathrm{Poisson}(\Lambda)\ \text{and}\ \Lambda \sim \mathrm{Gamma}(a_{\Lambda}, b_{\Lambda})\ \text{for}\ a_{\Lambda},b_{\Lambda}>0$.
\begin{prop}\label{prop_M_na}
For the full conditional distribution of $M_{na}$, the inequalities
\begin{align}
    \mathbb{P}(M_{na} \geq 1\mid U_{n}=u,\ k,\ \Lambda) &\leq \Lambda\psi(u)\left(1 + \frac{1}{\Lambda\psi(u) + k}\right),\label{M_na_1}\\
    \mathbb{P}(M_{na} \geq 1\mid U_{n}=u,\ k) &\leq \psi(u)\frac{a_{\Lambda}}{b_{\Lambda}}\left(1 + \frac{1}{k}\right)\label{M_na_2}
\end{align}
hold, where $\psi$ is the Laplace transform of the density $h$.
\end{prop}
The proof of the proposition is given in the Supplementary Material. Proposition \ref{prop_M_na} shows that $\psi(u)$ plays an important role in generating empty components. The critical difference between the inverse Gaussian and gamma distributions is the Laplace transform. Let $\psi_{\text{Gamma}}(u)$ be the Laplace transforms of $\text{Gamma}(\gamma, 1)$:
\begin{align}
    &\psi_{\text{Gamma}}(u) = \left(\frac{1}{1+u}\right)^\gamma,\quad u \geq 0.\label{Gamma_psi}
\end{align}
Laplace transforms $\psi_{\text{IGau}}$ and $\psi_{\text{Gamma}}$ are decreasing functions with respect to $u$. The former has exponential decay, while the latter is polynomial. Figure \ref{fig:laplace} shows graphs of the Laplace transforms of the inverse Gaussian and gamma distributions when the shape parameters are $\alpha,\gamma=1.0, 0.2, 10^{-1},10^{-2},10^{-3}$ and the scale parameters are $1$. It can be seen that $\psi_{\text{IGau}}$ decreases much faster than $\psi_{\text{Gamma}}$, contributing to the suppression of $M_{na}$.

Inequalities \eqref{M_na_1} and \eqref{M_na_2} are not for marginal posterior distributions, but for full conditional distributions of $M_{na}$. This is due to the fact that the marginal posterior distributions of $U_{n}$ are analytically intractable.

\subsection{Relationship with the telescoping sampling}
In this section, we clarify the connection between the blocked Gibbs sampling and the telescoping sampling. The blocked Gibbs sampling is derived from the posterior distribution of a random measure induced by the Norm-IFPP, whereas the telescoping sampling is a partially marginalized algorithm constructed from the EPPF. Although the two methods arise from fundamentally different formulations, they share several similarities, including the sampling of $M$. To the best of our knowledge, their relationship has not been clarified so far, as both methods were proposed independently around the same time. 

First, we explain that the blocked Gibbs sampling can be derived in a manner similar to the telescoping sampling scheme. Specifically, it can be derived from both the full mixture posterior $p(M, S, U_n, \tau_{1:M}, c \mid y) $ and the collapsed mixture posterior $p(M, \mathcal{C}, U_n, \tau_{1:k} \mid y)$, where $\mathcal{C}=\{\mathcal{C}_1,\dots,\mathcal{C}_k\}$ is a partition of $\{1,\dots,n\}$ into $k$ blocks, where $\mid\mathcal{C}_i\mid  = n_i$ for $i=1,\dots,k$ and $\sum_{i=1}^k n_i = n$, and the hyperparameters $\rho$ and $\eta$ are omitted for simplicity. This full mixture posterior is used to sample $\tau_{1:M}$, $c$, $S$, and $U_n$ (Steps~1,~2,~5,~6, and 7 in Algorithm~1). The collapsed mixture posterior is obtained by marginalizing out $S$, the unallocated parameters, 
and all assignments $c$ that induce the same partition $\mathcal{C}$ in the full mixture posterior. Thus, by Theorem~4.1 in \cite{argiento2022infinity}, the conditional distribution of $M$ given $U_n=u, \mathcal{C}$ is 
\begin{align*}
    \mathbb{P}(M=m \mid \mathcal{C}, U_n=u) 
    \propto \frac{m!}{(m-k)!} \psi(u)^{m-k} q_M(m),\quad m=k,k+1,\dots.
\end{align*}
This full conditional distribution of $M$ is equivalent to the equation \eqref{cond_M_na}, shifted by $k$. 

The latent variable $U_n$ is an essential distinction between the two methods, while the blocked Gibbs sampling shares many advantages with the telescoping sampling discussed in \cite{fruhwirth2021generalized}. In other words, the blocked Gibbs sampling can be regarded as a telescoping sampling augmented with the latent variable $U_n$. By this latent variable, the full conditional distribution $M_{na}$ (or $M$) can be easily sampled when $q_M$ is the Poisson distribution or negative binomial distribution. This is a major advantage over the telescoping sampling, because, in practice, an upper bound $M_{\mathrm{max}}$ is required and $M$ is sampled from a multinomial distribution over $\{k,\dots, M_{\mathrm{max}}\}$ due to the complex form of $M$'s full conditional distribution. 

\subsection{Extended telescoping sampling}
In this section, we extend the telescoping sampling so that it can be applied to model \eqref{eq:GMFM}, thereby relaxing the Dirichlet restriction in \cite{fruhwirth2021generalized}.

The conditional distributions for $\mathcal{C}$, given $M=m$ and $U_n =u$, and for $M$, given $\mathcal{C}, U_n=u$ are required to construct our algorithm under \eqref{eq:GMFM}.

\begin{prop}\label{prop:dynamic}
    Let $\mathcal{C} =\{\mathcal{C}_1,\dots,\mathcal{C}_k\}$ be a set partition with $n_j=| \mathcal{C}_j|$ and $ \sum_{j=1}^kn_j = n$. Under \eqref{eq:GMFM}, the EPPF is the following:
    \begin{align}\label{eq:EPPF_dynamic}
        p(\mathcal{C}) = \int_0^\infty \frac{u^{n-1}}{\Gamma(n)}\sum_{m=k}^\infty \frac{m!}{(m-k)!}\psi(u;m)^{m-k}\left\{\prod_{j=1}^k\kappa(u;n_j,m)\right\}q_M(m)\mathrm{d}u,\quad 
    \end{align}
    where $\psi(u;m)$ and $\kappa(u;n,m)$ are the Laplace transform and cumulant of $h_M$, respectively. Moreover, the conditional probability of $M = m$, for $m=k,k+1,\dots,$ is proportional to the following:
\begin{align}\label{eq:mgprop_Mna_dynamic}
        \mathbb{P}(M=m\mid U_n=u, \mathcal{C}) \propto \frac{m!}{(m-k)!}\psi(u;m)^{m-k}\left\{\prod_{j=1}^k\kappa(u;n_j,m)\right\}q_M(m).
    \end{align}
\end{prop}

From Proposition~\ref{prop:dynamic}, Algorithm~\ref{alg:ETS} is immediately derived, where only the parts that differ from Algorithm~1 are shown. If $h_M$ is $\text{Gamma}(\gamma/M, 1)$ or $\text{IGau}(\alpha/M, 1)$, 
Algorithm~\ref{alg:ETS} can be applied to the DMFM based on the Dirichlet and normalized inverse Gaussian distributions, respectively. The main difference from Algorithm~\ref{alg} is that the Laplace transform and the cumulant of $h_M$ depend on $M$. 
Since the full conditional distributions of $M$ (or $M_{na}$) and the summation $\sum_{m=k}^\infty \frac{m!}{(m-k)!}\left\{\prod_{j=1}^k\kappa(u;n_j,m)\right\}\psi(u;n_j, m)^{m-k}q_M(m)$
strongly depend on $\psi(u; m)$ and $\kappa(u; n, m)$, 
they cannot, in general, be expressed in a simple form. 
Thus, Step~3 in Algorithm~\ref{alg:ETS} is replaced by sampling from the full mixture posterior, 
and $M$ is not marginalized out in Step~3', as in \cite{fruhwirth2021generalized}. 
In addition, in Step~4, it is practically necessary to set an upper bound $M_{\text{max}}$ and sample from a multinomial distribution.
\begin{algorithm}[t]
\caption{Extended telescoping sampling}
\label{alg:ETS}
\begin{algorithmic}[htbp]
\STATE\textbf{Step 3.}\ Sample the hyperparameter $\eta$ of the $q_{M}$ from 
\[
p(\eta\mid\mathrm{rest}) \propto p(\eta)q_{M}(m).
\]
\STATE\textbf{Step 3'.}\ Sample the hyperparameter $\rho$ of the $h$ from 
\[
p(\rho\mid\mathrm{rest}) \propto \left\{\prod_{j=1}^k\kappa(u;n_j,m)\right\}\psi(u;m)^{m-k}p(\rho).
\]
\STATE\textbf{Step 4.}\ Sample $M$ from the following discrete distribution:
\[
\mathbb{P}(M=m\mid\mathrm{rest}) \propto \frac{m!}{(m-k)!}\left\{\prod_{j=1}^{k}\kappa(u;n_j,m)\right\}\psi(u;m)^{m-k}q_M(m),\quad m=k,k+1,\dots.
\]
\end{algorithmic}
\end{algorithm}

\section{Empirical demonstrations}
\label{sec:4}
We evaluate the performance of the MFM-IGau and MFM-Ga methods through some numerical experiments. Recall that $\gamma$ is the shape parameter of the prior distribution of unnormalized weights in the proposed method (MFM-IGau), while $\alpha$ is that of MFM-Ga.

\subsection{Clustering performance}
\label{sec:4.1}
In this subsection, we compare the clustering performance of the proposed methods, MFM-IGau and DMFM-IGau, with MFM-Ga and DMFM-Ga, using artificial and real data.

\subsubsection{Artificial data}
\label{sec:4.1.1}
In this simulation, we assume that $M_{\mathrm{true}} = k_{\mathrm{true}} = 8$, and data are generated from the following multivariate normal distribution mixture:
\begin{align*}
f(y\mid\mu_1,\dots,\mu_8) = \sum_{i=1}^{8} w_i N_{2}(y \mid \mu_i, \sigma^2I_2),
\end{align*}
where $w_i = i^{-1.75} / \sum_{i=1}^8 w_i$, $\mu_{1} = (2, 0)^{\top}, \mu_{2} = (2, 5)^{\top}, \mu_{3}=(6, 0)^{\top}, \mu_{4}=(6, 5)^{\top}, \mu_{5}=(10, 0)^{\top}, \mu_{6}=(10, 5)^{\top}, \mu_{7}=(14, 0)^{\top}, \mu_{8}=(14, 5)^{\top}$, $\sigma^2 = 0.45$, and $I_{2}$ is the $2\times 2$ identity matrix. We set $n=50, 100, 200, 300, 400$ and generate $50$ datasets for each $n$. 

This data generating process represents a situation in which many mixture components have small weights, so that not all components necessarily contribute to the observed data. When $n$ is small, it is difficult to obtain observations from all components, whereas as $n$ increases, it becomes more likely that observations are obtained from every component. 

MCMCs were run for 30,000 iterations, with the first 20,000 discarded as burn-in. We assume that $M-1\mid\Lambda \sim\text{Poisson}(\Lambda)$ and $\Lambda \sim \text{Gamma}(1, 0.1)$ and employ a multivariate normal kernel with the normal-inverse-Wishart hierarchical prior in Section~3.4.1, where $b_0 = \mathrm{median(y)}, B_0 = \mathrm{diag}(R_1^2,\dots,R_d^{2})$, $R_j$ is the range of data $y_{1:n}$ in the $j$th dimension, $c_0 = 2.5 + (d-1)/2, g_0 = 0.5 + (r-1)/2$, and $G_0 = 100 g_0/c_0\mathrm{diag}(1/R_1^2,\dots,1/R_d^2)$. Also, we set shape parameters $\alpha, \gamma = 1.0$ for MFM-IGau, DMFM-IGau, and MFM-Ga, and we assume a prior F distribution $F(6,3)$ to shape parameter DMFM-Ga as in \cite{fruhwirth2021generalized}. 

To measure performance, we consider the posterior mean of the adjusted Rand index (ARI) \citep{hubert1985comparing}, which is a measure of the goodness of clustering. When it is close to $1$, the assignment estimate is reasonable. The respective averages over $50$ repetitions are denoted by $\widehat{\mathrm{ARI}}$.

\begin{figure}[htbp]
    \centering
    \includegraphics[width=\linewidth]{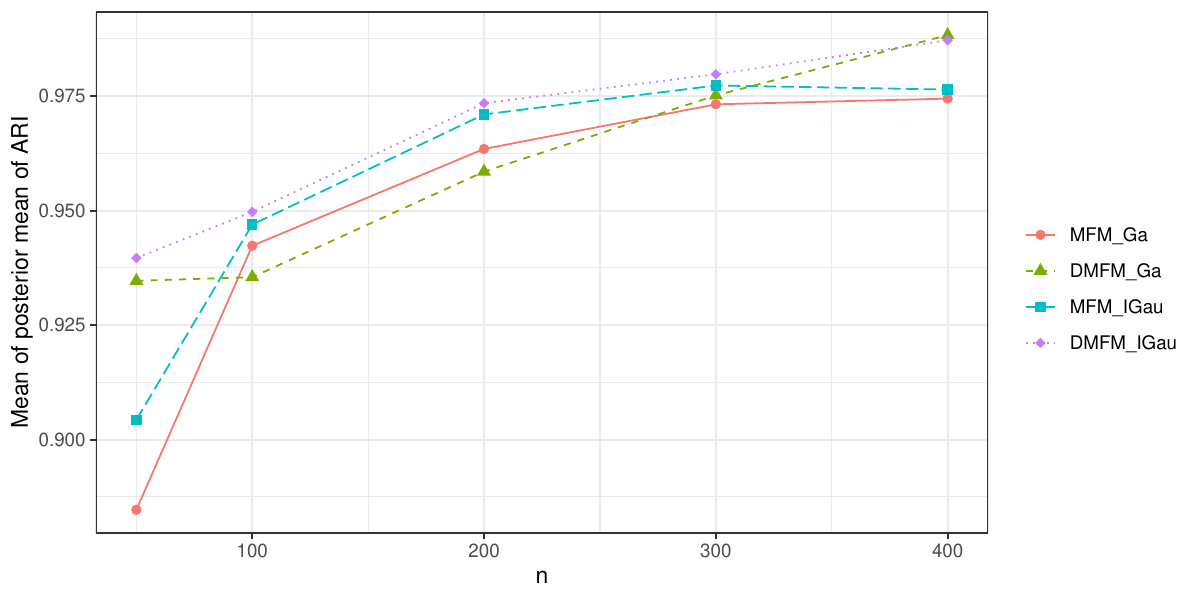}
    \caption{Line plots of the posterior mean of the ARI, averaged over $50$ Monte Carlo replications, for MFM-Ga, DMFM-Ga, MFM-IGau, and DMFM-IGau. The solid line with circles, dashed line with triangles, long-dashed line with squares, and dotted line with diamonds correspond to MFM-Ga, DMFM-Ga, MFM-IGau, and DMFM-IGau, respectively.}
    \label{fig:ARI_cluster_sim}
\end{figure}

Overall, figure~\ref{fig:ARI_cluster_sim} shows that $\widehat{\mathrm{ARI}}$ of all methods increase toward $1$ as $n$ becomes larger. MFM-IGau consistently attains higher clustering accuracy than MFM-Ga, and DMFM-IGau exhibits the best overall performance. Although the difference between DMFM-IGau and DMFM-Ga is small, DMFM-IGau does not incorporate learning for $\alpha$. This result indicates that stable posterior inference is achieved due to the less informative nature of the normalized inverse Gaussian distribution. The improvements in ARI appear modest, but even a small increase in ARI can be meaningful in this imbalance case. ARI differences can be structurally small and may not fully reflect improvements in detecting small clusters, since observations belonging to large weight components are typically identified easily. Therefore, even a modest gain in ARI may indicate better recovery of small clusters. Accurately detecting small clusters requires allocating non-negligible weight to them. In this regard, Figure 1 in the Supplementary Material shows that the normalized inverse Gaussian prior tends to allocate relatively larger weights to smaller components than the Dirichlet prior. Consequently, when the true weights are highly imbalanced, the normalized inverse Gaussian prior can facilitate the detection of small clusters.

\subsubsection{Thyroid data}
\label{sec:4.1.2}
We apply the proposed method MFM-IGau and DMFM-IGau to famous thyroid data. The main purpose of this analysis is to infer the number of clusters $k$ and the number of components $M$ and examine the impact of choice of $q_M$. The thyroid data are available from the R package {\tt mclust} and are well known as benchmark data for clustering. The sample size is $215$ and the dimension of the data is $6$. The thyroid disease of each patient is included and classified into three categories: normal, hypo, and hyper. The numbers of diseases are $150$, $30$, and $35$, respectively. Using these labels as true labels, the main interest is whether the number of components is estimated to be $3$. We use priors Poisson distribution $\mathrm{Poisson}(1)$, geometric distribution $\mathrm{Geo}(0.1)$, and beta negative binomial distribution $\mathrm{BNB}(1,4,3)$ for $M$, and $F$-distribution $F(6,3)$ for $\alpha$.

Table \ref{tab:thyroid_table} shows that the posterior modes for MFM-IGau and DMFM-IGau are all equal to $3$. These results are consistent with those obtained from DMFM-Ga. In addition, the table indicates that both the first and third quartiles are $3$, and that the posterior probabilities of $M_{na} = 0$ are very close to 1. This implies that the difference between the posterior number of components and the posterior number of clusters obtained by MFM-IGau and DMFM-IGau is extremely small. Furthermore, the interquartile range of $M$ is tighter than that reported in \cite{fruhwirth2021generalized}, providing stronger evidence that $M = 3$ under our proposed methods than under DMFM-Ga. In addition, the overall results do not depend on the choice of the prior $q_M$, as is the case for DMFM-Ga.

\begin{table}[htbp]
    \caption{Posterior summaries of $k$ and $M$ for the thyroid data. 
The leftmost values represent the posterior modes, the values in brackets denote the first and third quartiles, 
and the rightmost values correspond to the posterior probabilities of $M_{na}=0$.}
    \begin{center}
    \begin{tabular}{lccc} 
     \multicolumn{4}{c}{$p(k \mid \mathrm{data})$} \\ \hline
    & $\mathrm{Poi}(1)$ & $\mathrm{Geo}(0.1)$ & $\mathrm{BNB}(1,4,3)$ \\ \hline
    MFM-IGau & $3$\ $[3,3]$ & $3$\ $[3,3]$ & $3$\ $[3,3]$ \\ \hline
    DMFM-IGau & $3$\ $[3,3]$ & $3$\ $[3,3]$ & $3$\ $[3,3]$ \\ \hline
    &&& \\
    \multicolumn{4}{c}{$p(M\mid \mathrm{data})$} \\ \hline
    & $\mathrm{Poi}(1)$ & $\mathrm{Geo}(0.1)$ & $\mathrm{BNB}(1,4,3)$ \\ \hline
    MFM-IGau & $3$\ $[3,3]$\ $0.999$ & $3$\ $[3,3]$\ $0.998$ & $3$\ $[3,3]$\ $0.999$ \\ \hline
    DMFM-IGau & $3$\ $[3,3]$\ $0.997$ & $3$\ $[3,3]$\ $0.999$ & $3$\ $[3,3]$\ $0.992$ \\\hline
\end{tabular}
\end{center}
\label{tab:thyroid_table}
\end{table}

\subsection{Density estimation}
\label{sec:4.2}
In this section, we compare MFM-IGau and MFM-Ga using the Galaxy data. The aim of this analysis is to examine how sensitive the posterior distribution of $k$ and the density estimation based on the predictive distribution are to the choice of shape parameters.

We used the famous Galaxy dataset, which is a small dataset consisting of $82$ velocities (km/sec) of different galaxies. The data are widely used in nonparametric Bayesian statistics as a benchmark for density estimation and cluster analysis. The details and a more comprehensive analysis of the data were given in \cite{roeder1990density} and \cite{grun2022many}.

We set $M-1\mid\Lambda \sim \text{Poisson}(\Lambda)$ and $\Lambda \sim \text{Ga}(1, 1/5)$. We also employ the univariate normal kernel $N(y\mid\mu, \sigma^2)$, and the normal-inverse gamma conjugate prior in Section~3.4.1, $N(\mu\mid m_{0}, \sigma^2/\tau) \times \text{IG}(\sigma^2\mid c_{0}, C_{0})$, as the prior of the parameters in the kernel. Moreover, we assume that the prior of $C_{0}$ is $\text{Gamma}(d_{0}, D_{0})$, and we set $m_{0}=(\max(\mathrm{data})+\min(\mathrm{data}))/2,\ d_{0}=0.2,\ D_{0}=10/(\max(\mathrm{data})+\min(\mathrm{data}))^2$ as in \cite{richardson1997bayesian}.
The parameter $\tau$ controls the smoothness of the estimated density function. We assume that the prior of the smoothing parameter $\tau$ is $\text{Gamma}(w, W)$, where $w=0.5$ and $W=50$ as in \cite{escobar1995bayesian}. Parameters $\tau$ and $C_{0}$ should be carefully learned from the data, since density estimation is sensitive to their choice. MCMC was run for 100,000 iterations with a burn-in of 90,000. Each value of $\alpha=1.0, 0.2, 10^{-1}, 10^{-2}, 10^{-3}$ 
is considered, and $\gamma$ is set as $\gamma=\alpha$ in each case. Note that the first and second moments of $\mathrm{IGau}(\alpha, 1)$ and $\mathrm{Gamma}(\gamma, 1)$ are both equal to $\alpha$ and $\gamma$, respectively; thus, this corresponds to a moment-matching setting. We evaluated the result of density estimation and posterior probabilities of $k$ for each shape.

\begin{figure}[tbp]
    \centering
    \includegraphics[width=\linewidth]{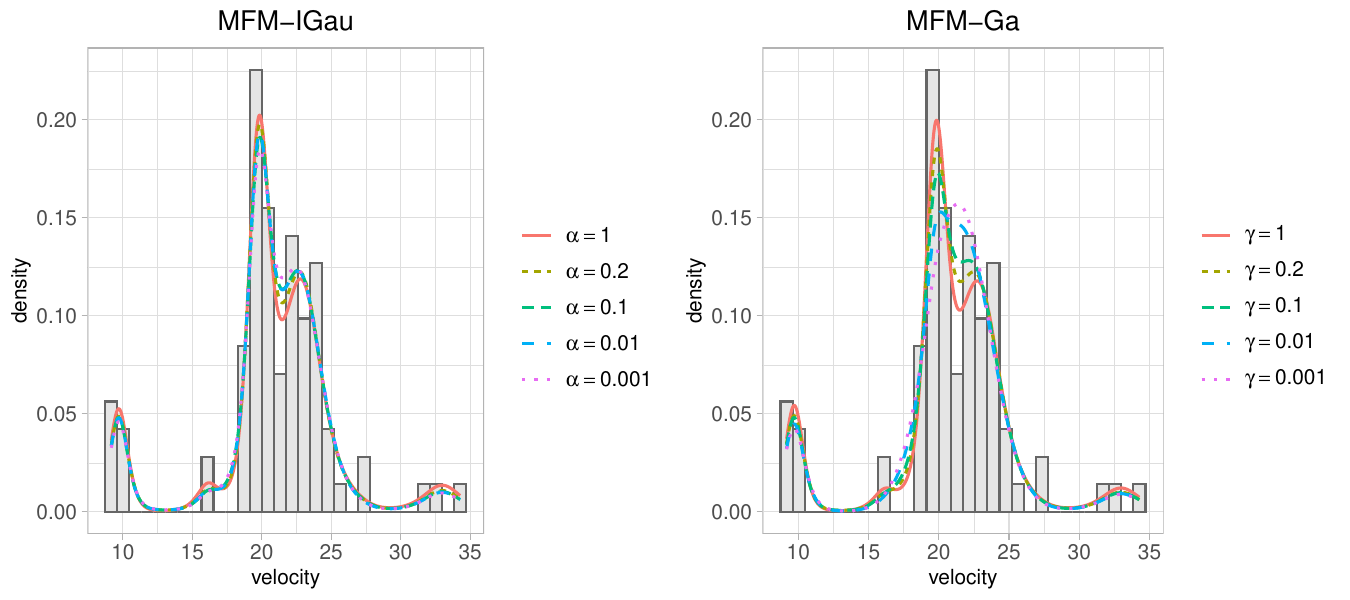}
    \caption{Results of density estimation for the Galaxy data using MFM-IGau (left) and MFM-Ga (right).}
    \label{fig:galaxy_densities}
\end{figure}

We show the estimated density functions using the posterior means in Figure \ref{fig:galaxy_densities}. From the figure, it is observed that the shapes of the estimated densities using MFM-IGau do not depend on the choice of the shape parameter $\alpha$. However, results using the MFM-Ga method seem to be strongly influenced by the choice of the shape parameter $\gamma$. For $\gamma=10^{-2}$ and $10^{-3}$, MFM-Ga cannot capture two large peaks in the middle.

The number of clusters in the Galaxy data has been reported as $5$ or $6$ in existing studies. From Table \ref{tab:M_k_table_galaxy}, in MFM-IGau, the posterior distributions of $k$ have large probabilities at $k=5$ and $6$ for all $\alpha$, and the posterior distributions of $k$ under MFM-IGau are more similar to each other across different values of $\alpha$ than those under MFM-Ga. Focusing on the number of clusters $k$, it is interesting that MFM-IGau is more robust than MFM-Ga with respect to the choice of the shape parameter in estimating $k$. This suggests that the prior distribution of $k$ based on MFM-IGau is less informative than MFM-Ga, i.e., the same relationship holds for MFM-IGau and MFM-Ga as for the normalized inverse Gaussian process and Dirichlet process. 
\begin{table}[tbp]
\caption{Posterior probabilities of the number of clusters $k$ for the Galaxy data.}
    \label{tab:M_k_table_galaxy}
    \begin{center}
    \begin{tabular}{lcccccccc}
    \multicolumn{9}{c}{MFM-IGau} \\ \hline
               & $k \leq 3$ & $k=4$ & $k=5$ & $k=6$ & $k=7$ & $k=8$ & $k=9$ & $k\geq10$ \\ \hline
  $\alpha=1.0$  & $0.000$ & $0.092$ & $0.241$ & $0.331$ & $0.164$ & $0.092$ & $0.046$ & $0.034$ \\ \hline
    $\alpha=0.2$  & $0.156$ & $0.215$ & $0.206$ & $0.182$ & $0.118$ & $0.063$ & $0.033$ & $0.026$
    \\ \hline
    $\alpha=10^{-1}$  & $0.089$ & $0.127$ & $0.214$ & $0.224$ & $0.169$ & $0.094$ & $0.052$ & $0.031$
    \\ \hline
    $\alpha=10^{-2}$  & $0.088$ & $0.123$ & $0.201$ & $0.232$ & $0.160$ & $0.107$ & $0.054$ & $0.035$
    \\ \hline
    $\alpha=10^{-3}$  & $0.082$ & $0.152$ & $0.256$ & $0.221$ & $0.143$ & $0.081$ & $0.039$ & $0.025$
    \\ \hline
   &&&&&&\\
    \multicolumn{9}{c}{MFM-Ga} \\ \hline
               & $k \leq 3$ & $k=4$ & $k=5$ & $k=6$ & $k=7$ & $k=8$ & $k=9$ & $k\geq10$ \\ \hline
  $\gamma=1.0$  & $0.044$ & $0.120$ & $0.232$ & $0.236$ & $0.188$ & $0.104$ & $0.044$ & $0.033$ \\ \hline
    $\gamma=0.2$  & $0.145$ & $0.198$ & $0.304$ & $0.186$ & $0.098$ & $0.040$ & $0.019$ & $0.009$
    \\ \hline
    $\gamma=10^{-1}$  & $0.145$ & $0.198$ & $0.304$ & $0.186$ & $0.098$ & $0.040$ & $0.019$ & $0.009$
    \\ \hline
    $\gamma=10^{-2}$  & $0.775$ & $0.195$ & $0.024$ & $0.005$ & $0.000$ & $0.000$ & $0.000$ & $0.000$
    \\ \hline
    $\gamma=10^{-3}$  & $0.983$ & $0.018$ & $0.000$ & $0.000$ & $0.000$ & $0.000$ & $0.000$ & $0.000$
    \\ \hline
\end{tabular}
\end{center}
\end{table}

The shape parameter of MFM-Ga should be chosen carefully because it has a significant impact on clustering, density estimation, and the appearance of empty components. However, MFM-IGau is much more robust than MFM-Ga with respect to the choice of the shape parameter. This is consistent with the earlier remark that the normalized inverse Gaussian distribution is less informative and more robust to parameter specification than the Dirichlet distribution, leading to more stable inference. Note that results do not differ substantially when information measures such as the expected Gini index are matched.

\subsection{Community detection}
\label{sec:4.3}
We apply the proposed method to community detection for network data. Similar comparisons as in Section \ref{sec:4.1} are made for both artificial and real data. Since the number of components of the finite mixture models is equivalent to the number of communities of the network, we use the same notation $M$ to denote the number of communities. 

\subsubsection{Artificial data}
\label{sec:4.3.1}
First, we illustrate the performance of the proposed method using simulation data. We assume that the true number of communities and clusters is $2$, denoted by $M_{\mathrm{true}} = k_{\text{true}}=2$, and the number of nodes in the network is set as $n=50$. We generated 50 datasets under this setting. We here consider the balanced network, in that the true allocation consists of $M_{\mathrm{true}}$ communities with $25$ nodes. For the true probability matrix $Q$, we assume that each component is expressed by $q_{rs} = q + (p - q)I(r=s)$, where $q = 0.1$, $p=0.35$, and $1 \leq r \leq s \leq M_{\mathrm{true}}$. 

We use MFM-Geng, which is suggested by \cite{geng2019probabilistic} and equivalent to MFM-Ga, and DMFM-Ga as baseline methods. In MFM-Geng, $\gamma$ is fixed at $1$, and $M$ is sampled by the post-preseeding step. We assign a $\text{BNB}(1,4,3)$ prior to $M-1$ and an $\mathcal{F}(6,3)$ prior to the shape parameters, and employ (\ref{bi_SBM}) as a kernel and a prior, where $a_{Q}=b_{Q}=1$.

We evaluate the performance in terms of $\widehat{\mathrm{ARI}}$, $\widehat{\mathbb{P}}(M_{na}=0)$, and the following criterias:
\begin{itemize}
\item Modularity (Mod) \citep{newman2004finding}: This index measures the quality of the estimated community structure in a network. We computed this value based on the MAP estimate of the label $c$ obtained by the method of \cite{dahl2009modal}. We reported $\widehat{\mathrm{Mod}}$, averaged over $50$ replications.
\item MCMC run times (seconds): All computations were implemented in R (version 4.4.0) on a laptop equipped with an Apple M2 chip (8-core CPU) and 16 GB of RAM, running macOS Sonoma 14.5. 
MCMCs were run for 100,000 iterations with a burn-in of 50,000 (no thinning).
\end{itemize} 

\begin{figure}[tbp]
    \centering
    \includegraphics[width=\linewidth, keepaspectratio]{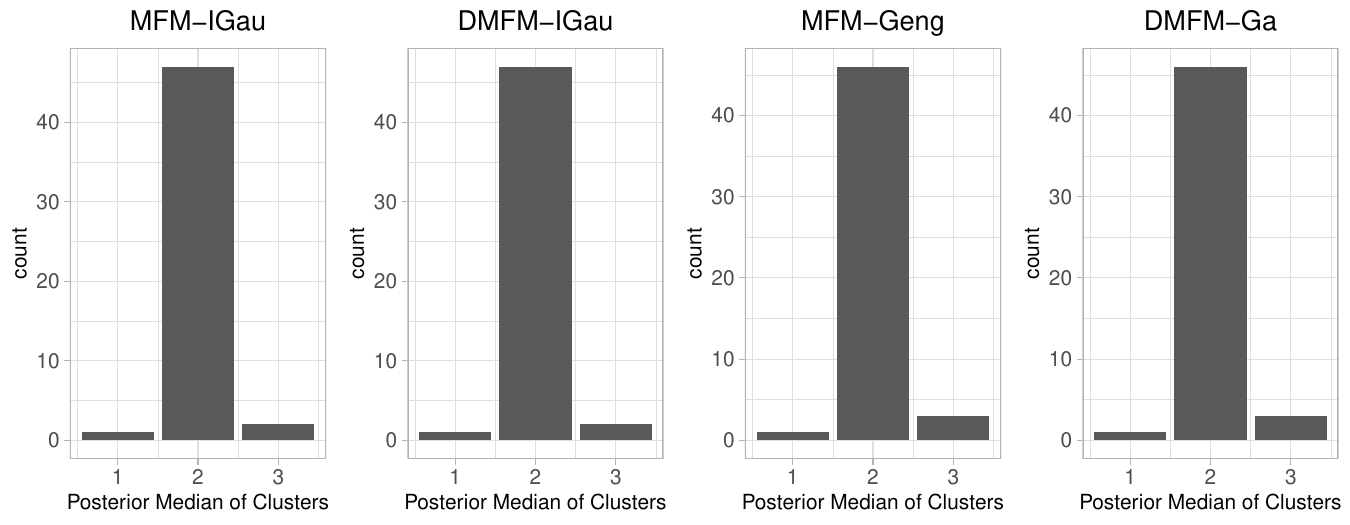}
    \caption{Bar plots of the posterior median of clusters $k$. From left to right, the models are MFM-IGau, DMFM-IGau, MFM-Geng, and DMFM-Ga.}
    \label{fig:bar_pm_clust}
\end{figure}
From Figure~\ref{fig:bar_pm_clust} and the $\widehat{\mathrm{ARI}}$ values in Table~\ref{tab:community_detection_table}, 
the posterior estimates of $k$ and the clustering accuracy are comparable across all methods. Although DMFM-Ga achieves a slightly higher ARI than the other methods, 
MFM-IGau and DMFM-IGau outperform MFM-Geng and DMFM-Ga in terms of modularity. This result indicates that MFM-IGau and DMFM-IGau better capture the community structures given the adjacency matrix. From the $\widehat{\mathbb{P}}(M_{na}=0)$ values in Table~\ref{tab:community_detection_table}, MFM-IGau and DMFM-IGau support $M = 2$ more tightly than the other methods. This result is also consistent with Proposition~\ref{prop_M_na}. In Table~\ref{tab:community_detection_table}, the most pronounced differences among the methods appear in terms of MCMC run time. MFM-IGau has the shortest run time, which can be attributed to the fact that all random numbers required at each step of the MCMC algorithm can be generated straightforwardly. 
DMFM-IGau and DMFM-Ga require more time than MFM-IGau, because sampling $M$ from a multinomial distribution with an upper bound is necessary. 
DMFM-IGau is slower than DMFM-Ga, reflecting the greater complexity of the cumulant of the inverse Gaussian distribution compared with that of the gamma distribution. The greater computational efficiency than that of MFM-Geng demonstrates the effectiveness of our proposed algorithms, since Algorithms~1 and~2 are used in MFM-IGau, DMFM-IGau and DMFM-Ga.  In Bayesian community detection, the computational bottleneck is typically the update of the label variables $c$. Under a restaurant-process update, labels are inherently updated sequentially, and each update requires case distinctions such as whether a node forms a singleton cluster or whether a new cluster is created. In addition, if a new cluster is created, the associated parameters must be updated immediately.  In contrast, under MFM-IGau (including DMFM-IGau and DMFM-Ga), the label update is equivalent in form to that of a finite mixture model with a fixed number of components: each node's assignment is evaluated against a fixed set of candidates using the same update rule. Therefore, each update can be implemented without such branching, and it suffices to perform component relabeling and the removal of empty components after completing the label updates. In this setting, the MFM-based methods outperformed other representative community detection methods, including LEM \citep{PhysRevE.74.036104}, HMM \citep{Blondel_2008}, and MH-MCMC \citep{PhysRevLett.117.078301}. The corresponding numerical results are reported in the Supplementary Material.
\begin{table}[tbp]
    \caption{Results of community detection over 50 simulated networks: $\widehat{\mathrm{ARI}}$, $\widehat{\mathbb{P}}(M_{na}=0)$, $\widehat{\mathrm{Mod}}$, and MCMC run time (seconds). Standard deviations are given in parentheses.}
    \begin{center}
    \begin{tabular}{lcccc} 
    \\ \hline
    & MFM-IGau & DMFM-IGau & MFM-Geng & DMFM-Ga \\ \hline
    \rule{0pt}{3.0ex} $\widehat{\mathrm{ARI}}$ & $0.853$ & $0.855$ & $0.857$ & $0.859$ \\
    & $(0.085)$ & $(0.084)$ & $(0.083)$ & $(0.083)$ \\\hline
    \rule{0pt}{3.0ex} $\widehat{\mathrm{Mod}}$ & $0.252$ & $0.258$ & $0.251$ & $0.253$ \\
& $(0.060)$ & $(0.052)$ & $(0.057)$ & $(0.058)$ \\\hline
\rule{0pt}{3.0ex}$\widehat{\mathbb{P}}(M_{na}=0)$ & $0.991$ & $0.973$ & $0.938$ & $0.823$ \\
& $(0.003)$ & $(0.007)$ & $(0.013)$ & $(0.027)$ \\\hline
\rule{0pt}{3.0ex}MCMC run time (sec) & $210.096$ & $305.166$ & $666.522$ & $268.927$ \\
& $(20.590)$ & $(30.355)$ & $(40.760)$ & $(26.479)$ \\\hline
\end{tabular}
\end{center}
\label{tab:community_detection_table}
\end{table}

\subsubsection{Dolphins' social network data}
\label{subsec:4.3.2}
A dolphins' social network is often used as a benchmark. Data can be obtained at \url{http://www-personal.umich.edu/mejn/netdata/}. The data constitute an undirected graph and express a small-scale animal social network with $64$ bottlenose dolphins off Doubtful Sound, New Zealand. Each node represents a dolphin, and an edge is drawn if two dolphins appear to be closely related to each other. From previous studies, it is well known that the network has two communities. The details of the data are found in \cite{lusseau2003bottlenose}.

We compare the posterior distributions of the numbers of clusters and communities among MFM-IGau, DMFM-IGau, MFM-Geng, and DMFM-Ga. In addition, we construct a co-clustering matrix to quantify the uncertainty of clustering and compute the modularity value, which is calculated based on the MAP estimator derived by the method of \cite{dahl2009modal}. Except for setting $a_Q = b_Q =3.0$, the model is identical to that used in Section~4.3.1.
\begin{table}[tbp]
    \caption{Posterior summaries of $k$, $M$, and modularity for the dolphin data. The leftmost values represent the posterior modes, the values in brackets denote the 5\% and 95\% quantiles, and the rightmost values correspond to the posterior probabilities of $M_{na}=0$.}
    \begin{center}
    \begin{tabular}{lcccc} 
     \hline
    & MFM-IGau & DMFM-IGau & MFM-Geng & DMFM-Ga\\ \hline
    $p(k\mid \mathrm{data})$ & $2$\ $[2,2]$ & $2$\ $[2,3]$ & $2$\ $[2,2]$ & $2$\ $[2,2]$\\ \hline
    $p(M\mid \mathrm{data})$ & $2$\ $[2,2]$\ $0.995$ & $2$\ $[2,2]$\ $0.984$& $2$\ $[2,3]$ $0.955$ & $2$\ $[2,3]$\ $0.851$ \\\hline
    modularity & $0.373$ & $0.385$ & $0.371$ & $0.370$ \\\hline
\end{tabular}
\end{center}
\label{tab:dolphin_com}
\end{table}
From Table \ref{tab:dolphin_com}, although all the results for the number of clusters $k$ are the same, the posterior distributions of $M$ obtained from MFM-IGau and DMFM-IGau support $M=2$ more strongly and more tightly than those obtained from MFM-Geng and DMFM-Ga. In addition, from the viewpoint of modularity, MFM-IGau and DMFM-IGau perform slightly better than MFM-Geng and DMFM-Ga. Finally, the co-clustering matrices obtained from MFM-IGau and DMFM-IGau are very similar to those from MFM-Geng, indicating that these methods produce reasonable clustering results.
\begin{figure}[htbp]
    \centering
    \includegraphics[width=\linewidth]{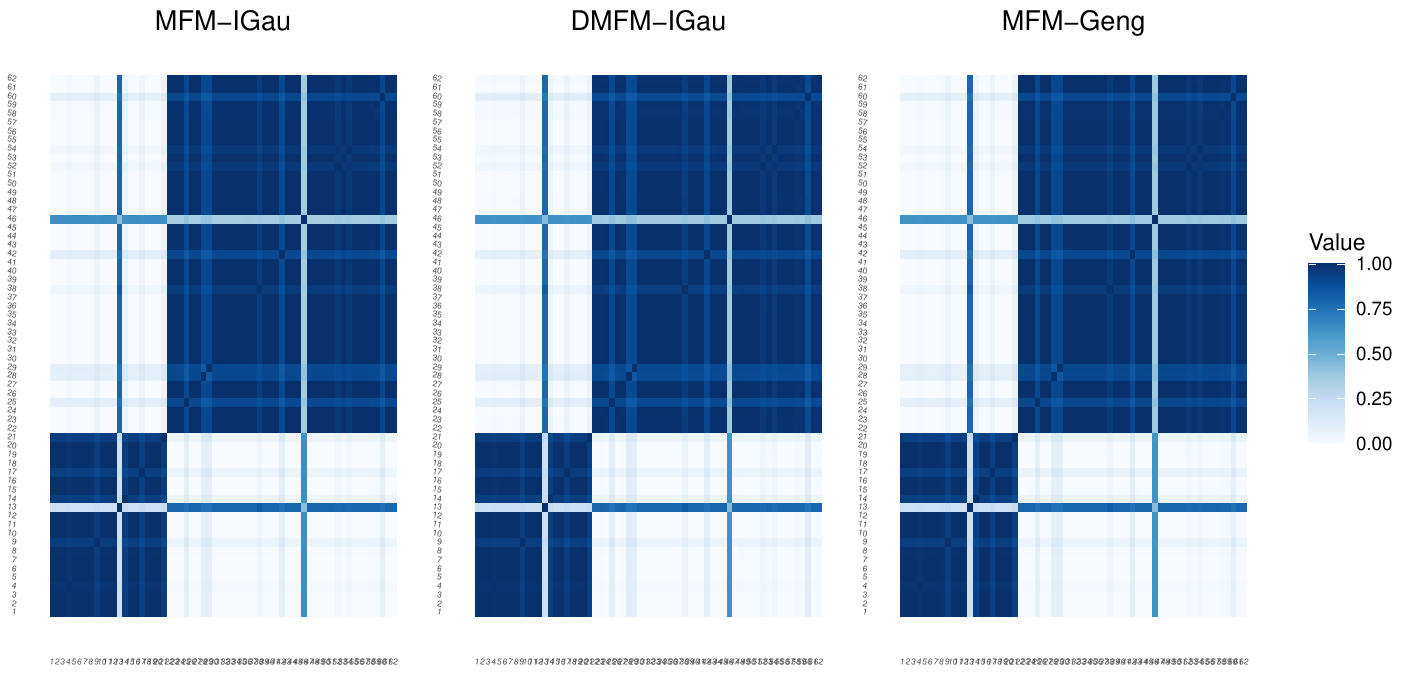}
    \caption{Co-clustering matrix with the MFM-IGau with $\alpha=1.0$ (left) and the MFM-Geng with $\gamma=1.0$ (right) for the dolphins' social network data.}
    \label{fig:heat_dol_MFM_Inv_Ga}
\end{figure}

\section{Concluding remarks}
\label{sec:5}
We proposed MFM and DMFM based on the normalized inverse Gaussian distribution and constructed an efficient posterior sampling algorithm based on \cite{argiento2022infinity} and \cite{fruhwirth2021generalized}. We illustrate the performance of the proposed method for clustering, density estimation, and community detection, compared to existing MFM and DMFM models based on the Dirichlet distribution. 

The proposed method has several limitations. Because \(U_n\) is difficult to marginalize out, some model quantities do not have closed-form marginal distributions. In particular, the model does not yield a tractable prior distribution for the number of clusters, making it difficult to incorporate interpretable subjective prior information \citep[see, e.g.,][]{zito2024bayesian}. In addition, MFM weights are obtained by normalizing independent random variables, which restricts the dependence structure among categories. Like the Dirichlet distribution, the normalized inverse Gaussian distribution induces negative covariances and may therefore be unsuitable for data exhibiting positive correlations, such as symbiotic proportions, disease complications, or gene expression data. A promising direction is to relax the independence assumption of the Norm-IFPP framework of \cite{argiento2022infinity} and construct MFMs with positively correlated weights. The proposed model could also be extended to spatial data \citep[e.g.,][]{geng2021bayesian}, functional data \citep[e.g.,][]{hu2023bayesian}, and stochastic block models with weighted edges or degree correction.

\section*{Acknowledgement}
 We thank Naoki Awaya for his valuable comments on an early revision of the manuscript. This work was supported by JST SPRING grant Number: JPMJSP2132, and partially supported by the Japan Society for the Promotion of Science grant number: 21K13835.

\bibliographystyle{chicago}
\bibliography{cite_arXiv}

\begin{thebibliography}{}

\bibitem[\protect\citeauthoryear{Argiento, Cremaschi, and Guglielmi}{Argiento et~al.}{2014}]{argiento2014density}
Argiento, R., A.~Cremaschi, and A.~Guglielmi (2014).
\newblock A “density-based” algorithm for cluster analysis using species sampling gaussian mixture models.
\newblock {\em Journal of Computational and Graphical Statistics\/}~{\em 23\/}(4), 1126--1142.

\bibitem[\protect\citeauthoryear{Argiento and Iorio}{Argiento and Iorio}{2022}]{argiento2022infinity}
Argiento, R. and M.~D. Iorio (2022).
\newblock {Is infinity that far? A Bayesian nonparametric perspective of finite mixture models}.
\newblock {\em {The Annals of Statistics}\/}~{\em 50\/}(5), pp. 2641--2663.

\bibitem[\protect\citeauthoryear{Blondel, Guillaume, Lambiotte, and Lefebvre}{Blondel et~al.}{2008}]{Blondel_2008}
Blondel, V.~D., J.-L. Guillaume, R.~Lambiotte, and E.~Lefebvre (2008, oct).
\newblock {Fast unfolding of communities in large networks}.
\newblock {\em Journal of Statistical Mechanics: Theory and Experiment\/}~{\em 2008\/}(10), P10008.

\bibitem[\protect\citeauthoryear{Dahl}{Dahl}{2009}]{dahl2009modal}
Dahl, D.~B. (2009).
\newblock Modal clustering in a class of product partition models.
\newblock {\em Bayesian Analysis\/}~{\em 4\/}(2), 243--264.

\bibitem[\protect\citeauthoryear{Escobar and West}{Escobar and West}{1995}]{escobar1995bayesian}
Escobar, M.~D. and M.~West (1995).
\newblock Bayesian density estimation and inference using mixtures.
\newblock {\em Journal of the American Statistical Association\/}~{\em 90\/}(430), 577--588.

\bibitem[\protect\citeauthoryear{Fr{\"u}hwirth-Schnatter}{Fr{\"u}hwirth-Schnatter}{2006}]{fruhwirth2006finite}
Fr{\"u}hwirth-Schnatter, S. (2006).
\newblock {\em Finite mixture and Markov switching models}.
\newblock Springer.

\bibitem[\protect\citeauthoryear{Fr{\"u}hwirth-Schnatter and Kaufmann}{Fr{\"u}hwirth-Schnatter and Kaufmann}{2008}]{frohwirth2008model}
Fr{\"u}hwirth-Schnatter, S. and S.~Kaufmann (2008).
\newblock Model-based clustering of multiple time series.
\newblock {\em Journal of Business \& Economic Statistics\/}~{\em 26\/}(1), 78--89.

\bibitem[\protect\citeauthoryear{Fr{\"u}hwirth-Schnatter, Malsiner-Walli, and Gr{\"u}n}{Fr{\"u}hwirth-Schnatter et~al.}{2021}]{fruhwirth2021generalized}
Fr{\"u}hwirth-Schnatter, S., G.~Malsiner-Walli, and B.~Gr{\"u}n (2021).
\newblock Generalized mixtures of finite mixtures and telescoping sampling.
\newblock {\em Bayesian Analysis\/}~{\em 16\/}(4), 1279--1307.

\bibitem[\protect\citeauthoryear{Fr{\"u}hwirth-Schnatter and Pyne}{Fr{\"u}hwirth-Schnatter and Pyne}{2010}]{fruhwirth2010bayesian}
Fr{\"u}hwirth-Schnatter, S. and S.~Pyne (2010).
\newblock Bayesian inference for finite mixtures of univariate and multivariate skew-normal and skew-t distributions.
\newblock {\em Biostatistics\/}~{\em 11\/}(2), 317--336.

\bibitem[\protect\citeauthoryear{Geng, Bhattacharya, and Pati}{Geng et~al.}{2019}]{geng2019probabilistic}
Geng, J., A.~Bhattacharya, and D.~Pati (2019).
\newblock Probabilistic community detection with unknown number of communities.
\newblock {\em Journal of the American Statistical Association\/}~{\em 114\/}(526), 893--905.

\bibitem[\protect\citeauthoryear{Geng, Shi, and Hu}{Geng et~al.}{2021}]{geng2021bayesian}
Geng, J., W.~Shi, and G.~Hu (2021).
\newblock {Bayesian nonparametric nonhomogeneous Poisson process with applications to USGS earthquake data}.
\newblock {\em Spatial Statistics\/}~{\em 41}, 100495.

\bibitem[\protect\citeauthoryear{Green}{Green}{1995}]{green1995reversible}
Green, P.~J. (1995).
\newblock {Reversible jump Markov chain Monte Carlo computation and Bayesian model determination}.
\newblock {\em Biometrika\/}~{\em 82\/}(4), 711--732.

\bibitem[\protect\citeauthoryear{Greve, Gr{\"u}n, Malsiner-Walli, and Fr{\"u}hwirth-Schnatter}{Greve et~al.}{2022}]{greve2022spying}
Greve, J., B.~Gr{\"u}n, G.~Malsiner-Walli, and S.~Fr{\"u}hwirth-Schnatter (2022).
\newblock Spying on the prior of the number of data clusters and the partition distribution in {Bayesian} cluster analysis.
\newblock {\em Australian \& New Zealand Journal of Statistics\/}~{\em 64\/}(2), 205--229.

\bibitem[\protect\citeauthoryear{Gr{\"u}n, Malsiner-Walli, and Fr{\"u}hwirth-Schnatter}{Gr{\"u}n et~al.}{2022}]{grun2022many}
Gr{\"u}n, B., G.~Malsiner-Walli, and S.~Fr{\"u}hwirth-Schnatter (2022).
\newblock How many data clusters are in the {Galaxy} data set? bayesian cluster analysis in action.
\newblock {\em Advances in data analysis and classification\/}~{\em 16\/}(2), 325--349.

\bibitem[\protect\citeauthoryear{Handcock, Raftery, and Tantrum}{Handcock et~al.}{2007}]{handcock2007model}
Handcock, M.~S., A.~E. Raftery, and J.~M. Tantrum (2007).
\newblock Model-based clustering for social networks.
\newblock {\em Journal of the Royal Statistical Society Series A: Statistics in Society\/}~{\em 170\/}(2), 301--354.

\bibitem[\protect\citeauthoryear{Holland, Laskey, and Leinhardt}{Holland et~al.}{1983}]{holland1983stochastic}
Holland, P.~W., K.~B. Laskey, and S.~Leinhardt (1983).
\newblock Stochastic blockmodels: First steps.
\newblock {\em Social networks\/}~{\em 5\/}(2), 109--137.

\bibitem[\protect\citeauthoryear{Hu, Geng, Xue, and Sang}{Hu et~al.}{2023}]{hu2023bayesian}
Hu, G., J.~Geng, Y.~Xue, and H.~Sang (2023).
\newblock Bayesian spatial homogeneity pursuit of functional data: an application to the us income distribution.
\newblock {\em Bayesian Analysis\/}~{\em 18\/}(2), 579--605.

\bibitem[\protect\citeauthoryear{Hubert and Arabie}{Hubert and Arabie}{1985}]{hubert1985comparing}
Hubert, L. and P.~Arabie (1985).
\newblock Comparing partitions.
\newblock {\em {Journal of Classification}\/}~{\em 2}, 193--218.

\bibitem[\protect\citeauthoryear{James, Lijoi, and Pr{\"u}nster}{James et~al.}{2009}]{james2009posterior}
James, L.~F., A.~Lijoi, and I.~Pr{\"u}nster (2009).
\newblock Posterior analysis for normalized random measures with independent increments.
\newblock {\em Scandinavian Journal of Statistics\/}~{\em 36\/}(1), 76--97.

\bibitem[\protect\citeauthoryear{Lijoi, Mena, and Pr{\"u}nster}{Lijoi et~al.}{2005}]{lijoi2005hierarchical}
Lijoi, A., R.~H. Mena, and I.~Pr{\"u}nster (2005).
\newblock Hierarchical mixture modeling with normalized inverse-{Gaussian} priors.
\newblock {\em Journal of the American Statistical Association\/}~{\em 100\/}(472), 1278--1291.

\bibitem[\protect\citeauthoryear{Lusseau, Schneider, Boisseau, Haase, Slooten, and Dawson}{Lusseau et~al.}{2003}]{lusseau2003bottlenose}
Lusseau, D., K.~Schneider, O.~J. Boisseau, P.~Haase, E.~Slooten, and S.~M. Dawson (2003).
\newblock The bottlenose dolphin community of doubtful sound features a large proportion of long-lasting associations: can geographic isolation explain this unique trait?
\newblock {\em Behavioral Ecology and Sociobiology\/}~{\em 54}, 396--405.

\bibitem[\protect\citeauthoryear{McLachlan, Bean, and Peel}{McLachlan et~al.}{2002}]{mclachlan2002mixture}
McLachlan, G.~J., R.~W. Bean, and D.~Peel (2002).
\newblock A mixture model-based approach to the clustering of microarray expression data.
\newblock {\em Bioinformatics\/}~{\em 18\/}(3), 413--422.

\bibitem[\protect\citeauthoryear{McLachlan, Lee, and Rathnayake}{McLachlan et~al.}{2019}]{mclachlan2019finite}
McLachlan, G.~J., S.~X. Lee, and S.~I. Rathnayake (2019).
\newblock Finite mixture models.
\newblock {\em Annual review of Statistics and its Application\/}~{\em 6\/}(1), 355--378.

\bibitem[\protect\citeauthoryear{Miller}{Miller}{2023}]{miller2023consistency}
Miller, J.~W. (2023).
\newblock Consistency of mixture models with a prior on the number of components.
\newblock {\em Dependence Modeling\/}~{\em 11\/}(1), 20220150.

\bibitem[\protect\citeauthoryear{Miller and Harrison}{Miller and Harrison}{2018}]{miller2018mixture}
Miller, J.~W. and M.~T. Harrison (2018).
\newblock Mixture models with a prior on the number of components.
\newblock {\em Journal of the American Statistical Association\/}~{\em 113\/}(521), 340--356.

\bibitem[\protect\citeauthoryear{Newman and Barkema}{Newman and Barkema}{1999}]{newman1999monte}
Newman, M.~E. and G.~T. Barkema (1999).
\newblock {\em Monte Carlo methods in statistical physics}.
\newblock Clarendon Press.

\bibitem[\protect\citeauthoryear{Newman and Girvan}{Newman and Girvan}{2004}]{newman2004finding}
Newman, M.~E. and M.~Girvan (2004).
\newblock Finding and evaluating community structure in networks.
\newblock {\em Physical Review E\/}~{\em 69\/}(2), 026113.

\bibitem[\protect\citeauthoryear{Newman}{Newman}{2006}]{PhysRevE.74.036104}
Newman, M. E.~J. (2006, Sep).
\newblock Finding community structure in networks using the eigenvectors of matrices.
\newblock {\em Phys. Rev. E\/}~{\em 74}, 036104.

\bibitem[\protect\citeauthoryear{Newman and Reinert}{Newman and Reinert}{2016}]{PhysRevLett.117.078301}
Newman, M. E.~J. and G.~Reinert (2016, Aug).
\newblock Estimating the number of communities in a network.
\newblock {\em Phys. Rev. Lett.\/}~{\em 117}, 078301.

\bibitem[\protect\citeauthoryear{Nobile}{Nobile}{1994}]{nobile1994bayesian}
Nobile, A. (1994).
\newblock {\em Bayesian analysis of finite mixture distributions}.
\newblock Carnegie Mellon University.

\bibitem[\protect\citeauthoryear{Nobile and Fearnside}{Nobile and Fearnside}{2007}]{nobile2007bayesian}
Nobile, A. and A.~T. Fearnside (2007).
\newblock Bayesian finite mixtures with an unknown number of components: The allocation sampler.
\newblock {\em Statistics and Computing\/}~{\em 17}, 147--162.

\bibitem[\protect\citeauthoryear{Richardson and Green}{Richardson and Green}{1997}]{richardson1997bayesian}
Richardson, S. and P.~J. Green (1997).
\newblock On {Bayesian} analysis of mixtures with an unknown number of components (with discussion).
\newblock {\em Journal of the Royal Statistical Society Series B: Statistical Methodology\/}~{\em 59\/}(4), 731--792.

\bibitem[\protect\citeauthoryear{Roeder}{Roeder}{1990}]{roeder1990density}
Roeder, K. (1990).
\newblock Density estimation with confidence sets exemplified by superclusters and voids in the galaxies.
\newblock {\em Journal of the American Statistical Association\/}~{\em 85\/}(411), 617--624.

\bibitem[\protect\citeauthoryear{Shi and Malik}{Shi and Malik}{2000}]{shi2000normalized}
Shi, J. and J.~Malik (2000).
\newblock Normalized cuts and image segmentation.
\newblock {\em IEEE Transactions on Pattern Analysis and Machine Intelligence\/}~{\em 22\/}(8), 888--905.

\bibitem[\protect\citeauthoryear{White and Smyth}{White and Smyth}{2005}]{white2005spectral}
White, S. and P.~Smyth (2005).
\newblock A spectral clustering approach to finding communities in graphs.
\newblock In {\em Proceedings of the 2005 SIAM international conference on data mining}, pp.\  274--285. SIAM.

\bibitem[\protect\citeauthoryear{Zito, Rigon, and Dunson}{Zito et~al.}{2024}]{zito2024bayesian}
Zito, A., T.~Rigon, and D.~B. Dunson (2024).
\newblock Bayesian nonparametric modeling of latent partitions via {Stirling}-gamma priors.
\newblock {\em Bayesian Analysis\/}~{\em 1\/}(1), 1--28.

\end{thebibliography}

\appendix
\section*{Appendix}
\addcontentsline{toc}{section}{Appendix}
\section{Proofs}
\subsection{Proof of the Cumulant of $\text{IGau}(\alpha, 1)$}
\begin{proof}
    We show by mathematical induction that the following equation holds.
\begin{align}
(-1)^n\frac{d^{n}}{du^{n}}\psi(u)
= \alpha^{n}\psi(u)\sum_{s=0}^{n-1}\frac{(n-1+s)!}{s!(n-1-s)!}(1 + 2u)^{-n/2-s/2}(2\alpha)^{-s},\label{nth_phi}
\end{align}
where $\psi(u) = \exp\left(\alpha\left(1 - \sqrt{1 + 2u}\right)\right), u \geq 0$. If $n=1$, the left side is $\alpha\psi(u)(1+2u)^{-1/2}$ and is equal to the right side. When $n=k$, we assume that \eqref{nth_phi} holds. We differentiate both sides of (\ref{nth_phi}) by u and $k+1$th derivative is 
\begin{align}
\alpha^{k+1}\psi(u)(1+2u)^{-k/2-1/2}\left(\sum_{s=0}^{k-1}A_{s}(1+2u)^{-s/2}(2\alpha)^{-s} + \sum_{s=0}^{k-1}B_{s}(1+2u)^{-s/2-1/2}(2\alpha)^{-s-1}\right),\notag
\end{align}
where
\[
A_{s} = \frac{(k-1+s)!}{s!(k-1-s)!},\quad B_{s}=\frac{2(k+s)!}{s!(k-1-s)!},\quad s=0,1,\dots,k-1.
\]
$B_{k-1}=(2k)!/k!$ and 
\begin{align}
    \sum_{s=0}^{k-1}B_{s}(1+2u)^{-s/2-1/2}(2\alpha)^{-s-1} &= \sum_{s=0}^{k-2}B_{s}(1+2u)^{-s/2-1/2}(2\alpha)^{-s-1} + \frac{(2k)!}{k!}(1+2u)^{-k/2}(2\alpha)^{-k}\notag\\
    &=\sum_{s=1}^{k-1}B_{s-1}(1+2u)^{-s/2}(2\alpha)^{-s} + \frac{(2k)!}{k!}(1+2u)^{-k/2}(2\alpha)^{-k}\notag.
\end{align}
For $s=1,\dots,k-1$, 
\[
A_{s} + B_{s-1} = \frac{(k+s)!}{s!(k-s)!}
\]
holds. Thus, 
\begin{align*}
&\sum_{s=0}^{k-1}A_{s}(1+2u)^{-s/2}(2\alpha)^{-s} + \sum_{s=0}^{k-1}B_{s}(1+2u)^{-s/2-1/2}(2\alpha)^{-s-1} \\
&=1 + \sum_{s=1}^{k-1}(A_{s} + B_{s-1})(1+2u)^{-s/2}(2\alpha)^{-s} + \frac{(2k)!}{k!}(1+2u)^{-k/2}(2\alpha)^{-k}\\
&=\sum_{s=0}^{k}\frac{(k+s)!}{s!(k-s)!}(1+2u)^{-s/2}(2\alpha)^{-s}.
\end{align*}
(\ref{nth_phi}) are derived by this equation, when $n=k+1$. Let $K_{\nu}(z)$ be a modified Bessel function of the second kind. In general,
\[
K_{n+1/2}(z) = K_{1/2}(z) \times \sum_{s=0}^{n}\frac{(n+s)!}{s!(n-s)!}(2z)^{-s}.
\]
Therefore, \eqref{nth_phi} holds.
\end{proof}

\subsection{Proof of Proposition~3.1}
\begin{proof}
    When $m \neq 0$, 
\[
  \mathbb{P}(M_{na}=m\mid U_{n}=u,\ k,\ \Lambda) = \frac{k}{\Lambda\psi(u) + k}\mathcal{P}_{0}(m;\Lambda\psi(u)) + \frac{\Lambda\psi(u)}{\Lambda\psi(u) + k}\mathcal{P}_{1}(m;\Lambda\psi(u)).
\]
Thus, the conditional expectation of $M_{na}$ is
\begin{align*}
      \mathbb{E}[M_{na}\mid U_{n}=u,\ k,\ \Lambda] &= \frac{k}{\Lambda\psi(u) + k} \times \Lambda\psi(u) + \frac{\Lambda\psi(u)}{\Lambda\psi(u) + k} \times ( \Lambda\psi(u) + 1 )\\
      &= \Lambda\psi(u)\left(1 + \frac{1}{\Lambda\psi(u) + k}\right).
  \end{align*}
  Furthermore,
  \begin{align*}
      \mathbb{E}[M_{na}\mid U_{n}=u,\ k] &= \mathbb{E}_{\Lambda}[\mathbb{E}[M_{na}\mid U_{n}=u,\ k,\ M=m,\ \Lambda]]\\
      &= \int \left(\Lambda\psi(u)\left(1 + \frac{1}{\Lambda\psi(u) + k}\right)\right)\frac{b_{\Lambda}^{a_{\Lambda}}}{\Gamma(a_{\Lambda})}\Lambda^{a_{\Lambda}-1}\exp(-b_{\Lambda}\Lambda)\mathrm{d}\Lambda\\
      &\leq \psi(u) \times \frac{a_{\Lambda}}{b_{\Lambda}} + \frac{\psi(u)}{k} \times \frac{a_{\Lambda}}{b_{\Lambda}}\\
      &= \psi(u) \times \frac{a_{\Lambda}}{b_{\Lambda}}\left(1 + \frac{1}{k}\right).
  \end{align*}
  Finally, the result follows from Markov's inequality.
\end{proof}

\subsection{Equivalence of Blocked Gibbs Sampler and Telescoping Sampler}
In this section, we show that the blocked Gibbs sampler is essentially equivalent to the telescoping sampler. 
Our strategy is to derive the computational algorithm under the framework of $(2.1)$ or $(2.4)$ in the same manner as the telescoping sampler. To this end, we require the EPPF induced by the Norm-IFPP, which is given below.
\begin{thm}[Theorem 4.1 in \cite{argiento2022infinity}]\label{Alg_EPPF}
    Let $\mathcal{C}=\{\mathcal{C}_1,\dots,\mathcal{C}_k\}$ be a partition of $\{1,\dots,n\}$ into $k$ blocks, where $|\mathcal{C}_i| = n_i$ for $i=1,\dots,k$ and $\sum_{i=1}^k n_i = n$. The EPPF $p(\mathcal{C})$ induced under $(2.1)$ or $(2.4)$ is
    \begin{align*}
    p(\mathcal{C}) &= \int_{0}^\infty \frac{u^{n-1}}{\Gamma(n)}\Psi(u;k)\left\{\prod_{j=1}^k\kappa(u;n_j)\right\}\mathrm{d}u,
    \end{align*}
    where
    \[
    \Psi(u;k) = \sum_{m=0}^\infty \frac{(m+k)!}{m!}\psi(u)^m q_M(m+k).
    \]
\end{thm}

Note that, for simplicity, we omit the hyperparameter $\eta$ of $q_M$ 
and the hyperparameter $\rho$ of $h$. The full mixture posterior $p(M, S, \tau_{1:M}, c \mid y)$ is 
\begin{align*}
    p(M, S, \tau, c \mid y) \propto
    \frac{1}{T^n}\left\{\prod_{m \in \mathcal{M}_{a}}S_{m}^{n_{m}}h(S_{m})f(y^{[n_m]} \mid \tau_{m})p_0(\tau_{m})\right\}\left\{\prod_{m \in \mathcal{M}_{na}}h(S_{m})p_0(\tau_{m})\right\}q_M(M),
\end{align*}
where $y^{[n_m]}$ denotes the $n_m$ observations belonging to the $m$th cluster $\mathcal{C}_m$. Also,
\[    
    U_{n}\mid T \sim \mathrm{Gamma}(n, T),\quad \frac{\Gamma(n)}{T^{n}} = \int_{0}^{\infty}e^{-Tu}u^{n-1}\mathrm{d}u.
\]
Using the data augmentation with $U_n$, we obtain the following:
\begin{align}
    p(M, S, U_n, \tau_{1:M}, c \mid y) 
    &\propto \frac{U_n^{n-1}}{\Gamma(n)}\left\{\prod_{m\in\mathcal{M}_a}f(y_j^{[n_m]} \mid \tau_{m})p_0(\tau_{m})\right\}\left\{\prod_{m\in\mathcal{M}_a}S_m^{n_m}\exp(-S_m U_n)h(S_m)\right\}\notag\\
    &\times\left\{\prod_{m\in\mathcal{M}_{na}}\exp(-S_m U_n)h(S_m) \right\}q_M(M).\notag
\end{align}
Marginalizing out the empty components, $S_1,\dots,S_M$, 
and all allocations $c$ that induce the same partition $\mathcal{C}$ yields the following expression for $p(M,\mathcal{C},U_n,\tau_{1:k}\mid y)$:
\begin{align}
    p(M, \mathcal{C}, U_n, \tau_{1:k} \mid y) \propto \left\{\prod_{m\in\mathcal{M}_a}f(y_j^{[n_m]} \mid \tau_{m})p_0(\tau_{m})\right\}
    p(\mathcal{C}, U_n\mid M)
    q_M(M). \notag
\end{align}
From Theorem~\ref{Alg_EPPF}, 
\begin{align}
    p(M, \mathcal{C}, U_n, \tau_{1:k} \mid y) \propto U_n^{n-1} \left\{\prod_{m\in\mathcal{M}_a}f(y_j^{[n_m]} \mid \tau_{m})p_0(\tau_{m})\right\}
    \frac{M!}{(M-k)!}\left\{\prod_{j=1}^k\kappa(n_j;U_n)\right\}\psi(U_n)^{M-k}q_M(M). \notag
\end{align}
Therefore, we obtain
\begin{align*}
    \mathbb{P}(M=m \mid \mathcal{C}, U_n=u) 
    \propto \frac{m!}{(m-k)!} \psi(u)^{m-k} q_M(m),\quad m=k,k+1,\dots.
\end{align*}

The extended telescoping sampler is immediately obtained in the same manner by applying Proposition~3.2.
\subsection{Proof of Proposition~3.2}
\begin{proof}
    This proposition follows from a slight modification of the proof of Theorem~\ref{Alg_EPPF}, with most steps carrying over directly. Let $\mathcal{A}(\mathcal{C})$ be the set of all labels $c$ that induces partition $\mathcal{C}$, and $c_1^\ast,\dots,c_k^\ast$ be unique values of $c$. The EPPF $p(\mathcal{C})$ is written by $p(\mathcal{C}) = \sum_{m=1}^\infty p(\mathcal{C}\mid M=m)q_M(m)$, since $q_M(0)=0$. Moreover,  
    \begin{align*}
        p(\mathcal{C}\mid M=m) 
        &= \mathbbm{1}_{\{k \leq m\}}\sum_{c \in \mathcal{A}(\mathcal{C})}p(c \mid M=m)\\
        &=\mathbbm{1}_{\{k \leq m\}}\sum_{c \in \mathcal{A}(\mathcal{C})}\int p(c \mid \pi, M=m)p(\pi\mid M=m)\mathrm{d}\pi\\
        &=\mathbbm{1}_{\{k \leq m\}}\sum_{c \in \mathcal{A}(\mathcal{C})}\mathbb{E}\left[\prod_{j=1}^k\pi_{c_j^\ast}^{n_j}\right].
    \end{align*}
    Note that $\pi_{c_j^\ast}^{n_j} = (S_{c_j^\ast}/T)^{n_j}$ and $\Gamma(n) = T^n = \int_0^\infty e^{-Tu}u^{n-1}\mathrm{d}u$. Thus,
    \begin{align*}
        \mathbb{E}\left[\prod_{j=1}^k\pi_{c_j^\ast}^{n_j}\right] 
        &= \mathbb{E}\left[\frac{1}{T^{n}}\prod_{j=1}^k S_{c_j^\ast}^{n_j}\right] \\
        &=\int_0^\infty \frac{u^{n-1}}{\Gamma(n)}\left\{\prod_{j=1}^k\int_0^\infty S_{c_j^\ast}^{n_j}e^{-S_{c_j^\ast}u}h_{M}(S_{c_j^\ast})\mathrm{d}S_{c_j^\ast}\right\}\left\{\prod_{t\notin(c_1^\ast,\dots,c_k^\ast)}\int_0^\infty e^{-S_t u}h_M(S_t)\mathrm{d}S_t\right\}\mathrm{d}u \\
        &=\int_0^\infty \frac{u^{n-1}}{\Gamma(n)}\left\{\prod_{j=1}^k\mathbb{E}\left[S_{c_j^\ast}^{n_j}e^{-S_{c_j^\ast}u}\right]\right\}\left\{\prod_{t\notin (c_1^\ast,\dots,c_k^\ast)}\mathbb{E}\left[e^{-S_t u}\right]\right\}\mathrm{d}u\\
        &=\int_0^\infty\frac{u^{n-1}}{\Gamma(n)}\left\{\prod_{j=1}^k\kappa(u;n_j,m)\right\}\psi(u;m)^{m-k}\mathrm{d}u,
    \end{align*}
    where, if $M=m$, $\psi(u;m) = \int_0^\infty e^{-uS}h_M(S)\mathrm{d}S$ and $\kappa(u;n,m) := (-1)^n \frac{d^n}{du^n}\psi(u;m) = \int_0^\infty S^ne^{-Su}h_M(S)\mathrm{d}u$. Here, if $M=m$, we have
    \[
    \#\mathcal{A}(\mathcal{C}) = \frac{m!}{(m-k)!k!}\times k! = \frac{k!}{(m-k)!}.
    \]
    Therefore,
    \begin{align*}
        p(\mathcal{C}) 
        &= \sum_{m=1}^\infty p(\mathcal{C}\mid M=m)q_M(m)\\
        &= \sum_{m=1}^\infty \mathbbm{1}_{\{k \leq m\}}\frac{m!}{(m-k)!}\int_0^\infty\left\{\prod_{j=1}^k\kappa(u;n_j,m)\right\}\psi(u;m)^{m-k}\mathrm{d}u\\
        &= \sum_{m=k}^\infty \frac{m!}{(m-k)!}\int_0^\infty\left\{\prod_{j=1}^k\kappa(u;n_j,m)\right\}\psi(u;m)^{m-k}\mathrm{d}u\\
        &= \int_0^\infty\frac{u^{n-1}}{\Gamma(n)}\Omega(u;k)\mathrm{d}u,
    \end{align*}
    where,
    \[
    \Omega(u;k) := \sum_{m=k}^\infty \frac{m!}{(m-k)!}\psi(u;m)^{m-k}\left\{\prod_{j=1}^k\kappa(u;n_j,m)\right\}q_M(m+k).
    \]
    Finally, the conditional distribution of $M$ is derived from the collapsed mixture posterior as follows:
    \begin{align*}
    p(M, \mathcal{C}, u, \tau_{1:k} \mid y) 
    &\propto \left\{\prod_{m\in\mathcal{M}_a}f(y_j^{[n_m]} \mid \tau_{m})p_0(\tau_{m})\right\}
    p(\mathcal{C}, u\mid M)
    q_M(M) \\
    &\propto\left\{\prod_{m\in\mathcal{M}_a}f(y_j^{[n_m]} \mid \tau_{m})p_0(\tau_{m})\right\}\frac{m!}{(m-k)!}\left\{\prod_{j=1}^k\kappa(u;n_j,m)\right\}\psi(u;m)^{m-k}q_M(M)\\
    \mathbb{P}(M=m\mid U_n=u, \mathcal{C}) &\propto \frac{m!}{(m-k)!}\psi(u;m)^{m-k}\left\{\prod_{j=1}^k\kappa(u;n_j,m)\right\}q_M(m).
\end{align*}
\end{proof}

\section{Additional Simulation Results}
\subsection{Gini Index Comparison}
In this section, we compare the Dirichlet distribution and the normalized inverse Gaussian distribution in terms of \emph{Gini index}. 

Let $x$ be a $d$-dimensional vector on $\mathbb{S}_d$. We define the Gini index of $x$ as
\[
G(x) = \sum_{i=1}^{d} x_i (1 - x_i).
\]
This measure indicates that when $G(x)$ is small, the point $x$ lies near a vertex or an edge of the simplex, whereas when $G(x)$ is large, the point $x$ lies near the barycenter of the simplex.

\begin{figure}[htbp]
    \centering
    \includegraphics[scale=0.8]{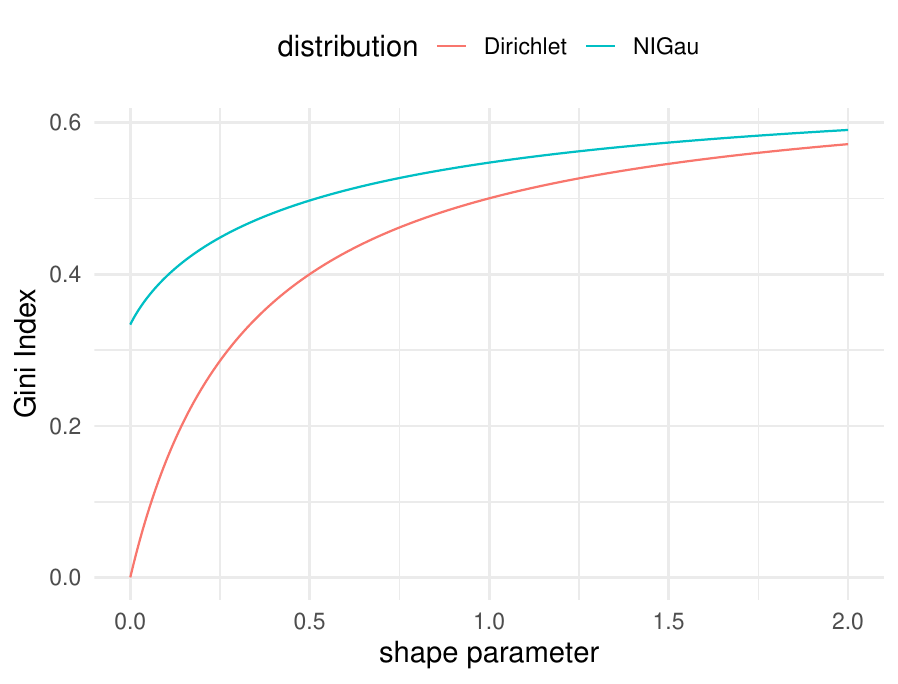}
    \caption{$\mathbb{E}[G(\pi)]$ for $\mathrm{Dirichlet}(\gamma,\gamma,\gamma)$ and $\mathrm{NIGau}(\alpha,\alpha,\alpha)$.  These values were computed on a grid of 10000 points between $0$ and $2$ (excluding $0$).}
    \label{fig:entropy_plot}
\end{figure}

We examine the expected values of the Gini index for (symmetry) Dirichlet and normalized inverse Gaussian distribution. Let $\pi=(\pi_1,\dots,\pi_d) \sim \text{NIGau}(\alpha,\dots,\alpha)$. From Proporition~2 in \cite{lijoi2005hierarchical}, 
\[
\mathbb{E}[\pi_i] = \frac{1}{d},\quad \text{Var}[\pi_i] = \alpha^2(d-1)e^{d\alpha}\Gamma(-2, d\alpha),\quad i=1,\dots,d,
\]
where $\Gamma(\cdot, \cdot)$ is the incomplete gamma function. Thus, we obtain immediately,
\[
\mathbb{E}[G(\pi)] = 1 - \frac{1}{d} -d\alpha^2(\alpha-1)e^{d\alpha}\Gamma(-2, d\alpha).
\]
From Figure~\ref{fig:entropy_plot}, values of the Dirichlet distribution decreases sharply as the parameter approaches the origin. In contrast, that of the normalized inverse Gaussian distribution decline much more gradually. This corresponds to the fact that, for the Dirichlet distribution with a small $\gamma$, the probability mass tends to concentrate on the vertices or edges of the simplex. In contrast, even when $\alpha$ is small, the normalized inverse Gaussian distribution does not induce the same sparsity as  the Dirichlet distribution, resulting in a more dispersed distribution over the simplex. Therefore, in settings such as the DMFM where the shape parameters naturally become small, using the normalized inverse Gaussian distribution can lead to more stable posterior inference.

\subsection{Estimating The Number of Components}
We conduct numerical experiments for MFM-IGau and MFM-Ga by matching moments and $\mathbb{E}[G(\pi)]$, focusing on the inference of the number of components. We assume that $M_{\mathrm{true}} = k_{\mathrm{true}} = 3$, and data are generated from the following:
\begin{align*}
f(y\mid \mu_1,\mu_2,\mu_3) = \sum_{i=1}^{3} w_i N_{2}(y \mid \mu_i, \sigma^2I_2),
\end{align*}
where $w_i = 1/3$, $\mu_{1} = (2, 0)^{\top}, \mu_{2} = (2, 5)^{\top}, \mu_{3}=(6, 0)^{\top}$, $\sigma^2 = 0.45$ and $I_{2}$ is the $2\times 2$ identity matrix. We set $n=300$ and generate $50$ datasets. We set $\alpha=0.1, 0.01, 0.001$ and consider the following two settings for $\gamma$:
\begin{itemize}
    \item Moment match: We set $\gamma=0.1,0.01,0.001$. Note that first and second moments of $\text{Gamma}(\gamma)$ and $\text{IGau}(\alpha, 1)$ are $\gamma$ and $\alpha$, respectively.
    \item $\mathbb{E}[G(\pi)]$ match: We set $\gamma=0.490, 0.352, 0.335$. For each $\alpha$, we compute $\mathbb{E}[G(\pi)]$ for $\text{NIGau}(\alpha,\alpha,\alpha)$ and determine the corresponding value of $\gamma$ for each $\alpha$, using the \texttt{uniroot} function in \textsf{R}. 
\end{itemize}
The remaining settings for kernels are the same as in Section 4.1.1. Each MCMC iteration is $30000$ and the first $20000$ samples are not used as a burn-in period. We assume that $M-1\mid \Lambda \sim\text{Poisson}(\Lambda)$ and $\Lambda \sim \text{Gamma}(1, 1)$.

In Table~\ref{tab:postprob_means_M}, we reported posterior probabilities averaged over $50$ repetitions. We observe that MFM-IGau consistently yields the highest (averaged) posterior probability at $M=3$. On the other hand, in the moment match case, MFM-Ga tends to overestimate $M$, because it produces many empty components when $\gamma$ is small. In the Gini index match case, MFM-Ga also assigns the highest posterior probability to $M=3$, similarly to MFM-IGau, but the probability is lower than that of MFM-IGau. Moreover, the results for very small $\gamma$, such as $\gamma = 0.1, 0.01, 0.001$, could also be interpreted from the perspective of the Gini index. When $\gamma$ is extremely small, the expected Gini index tends to decrease, and consequently the Dirichlet prior assigns very small probabilities to many components. Consequently, this prior structure may generate unnecessary empty components.
\begin{table}[htbp]
    \caption{Posterior probabilities and their standard deviations (shown in parentheses) of the number of components $M$ averaged over $50$ Monte Carlo replications. The highest value is bolded.}
    \begin{tabular}{lccccccccc}
   \multicolumn{10}{c}{MFM-IGau} \\ \hline
    & $M=1$ & $M=2$ & $M=3$ & $M=4$ & $M=5$ & $M=6$ &$M=7$ &$M=8$ &$M\geq9$  \\ \hline
   $\alpha=10^{-1}$ & $0.000$ & $0.000$ & $\bm{0.971}$ & $0.024$ & $0.004$ & $0.001$ & $0.000$ & $0.000$ & $0.000$ \\
   &$(0.000)$ & $(0.000)$ & $(0.059)$ & $(0.046)$ & $(0.021)$ 
&  $(0.003)$ & $(0.000)$ & $(0.000)$ & $(0.000)$ \\\hline
   $\alpha=10^{-2}$ & $0.000$ & $0.000$ & $\bm{0.959}$ & $0.036$ & $0.004$ & $0.001$ & $0.000$ & $0.000$ & $0.000$\\
   & $(0.000)$ & $(0.000)$ & $(0.063)$ & $(0.054)$ & $(0.008)$ & $(0.004)$ & $(0.002)$ & $(0.001)$ & $(0.000)$  \\\hline
   $\alpha=10^{-3}$ & $0.000$ & $0.000$ & $\bm{0.941}$ & $0.048$ & $0.007$ & $0.002$ & $0.001$ & $0.000$ & $0.001$ \\
   & $(0.000)$ & $(0.000)$ & $(0.111)$ & $(0.096)$ & $(0.014)$ & $(0.006)$ & $(0.003)$ & $(0.002)$ & $(0.003)$ \\\hline
   &&&&&&\\
   \multicolumn{10}{c}{MFM-Ga} \\ \hline
  $\gamma=10^{-1}$  & $0.000$ & $0.000$ & $\bm{0.331}$ & $0.301$ & $0.187$ & $0.098$ & $0.047$ & $0.021$ & $0.016$ \\ 
  & $(0.000)$ & $(0.000)$ & $(0.100)$ & $(0.021)$ & $(0.035)$ & $(0.028)$ & $(0.017)$ & $(0.009)$ & $(0.009)$ \\ \hline
   $\gamma=10^{-2}$  & $0.000$ & $0.000$ & $0.098$ & $0.172$ & $\bm{0.187}$ & $0.165$ & $0.128$ & $0.091$ & $0.159$\\ 
   & $(0.000)$ & $(0.000)$ & $(0.021)$ & $(0.024)$ & $(0.013)$ & $(0.005)$ & $(0.009)$ & $(0.012)$ & $(0.037)$ \\ \hline
   $\gamma=10^{-3}$  & $0.000$ & $0.000$ & $0.066$ & $0.131$ & $0.161$ & $0.159$ & $0.136$ & $0.107$ & $\bm{0.240}$\\ 
   & $(0.000)$ & $(0.000)$ & $(0.003)$ & $(0.004)$ & $(0.004)$ & $(0.004)$ & $(0.003)$ & $(0.003)$ & $(0.009)$ \\ \hline
   $\gamma=0.490$ & $0.000$ & $0.000$ & $\bm{0.798}$ & $0.170$ & $0.028$ & $0.004$ & $0.001$ & $0.000$ & $0.000$\\ 
   & $(0.000)$ & $(0.000)$ & $(0.070)$ & $(0.048)$ & $(0.019)$ & $(0.005)$ & $(0.001)$ & $(0.000)$ & $(0.000)$ \\ \hline
   $\gamma=0.352$  & $0.000$ & $0.000$ & $\bm{0.717}$ & $0.221$ & $0.050$ & $0.010$ & $0.002$ & $0.000$ & $0.000$\\ 
   & $(0.000)$ & $(0.000)$ & $(0.040)$ & $(0.024)$ & $(0.012)$ & $(0.004)$ & $(0.001)$ & $(0.000)$ & $(0.000)$\\\hline
   $\gamma=0.335$  & $0.000$ & $0.000$ & $\bm{0.693}$ & $0.237$ & $0.056$ & $0.012$ & $0.002$ & $0.000$ & $0.000$\\ 
   & $(0.000)$ & $(0.000)$ & $(0.053)$ & $(0.030)$ & $(0.016)$ & $(0.005)$ & $(0.001)$ & $(0.000)$ & $(0.000)$\\\hline
   &&&&&&\\
\end{tabular}
\label{tab:postprob_means_M}
\end{table}

\subsection{Community Detection}
\begin{figure}[htbp]
    \centering
    \includegraphics[width=\linewidth]{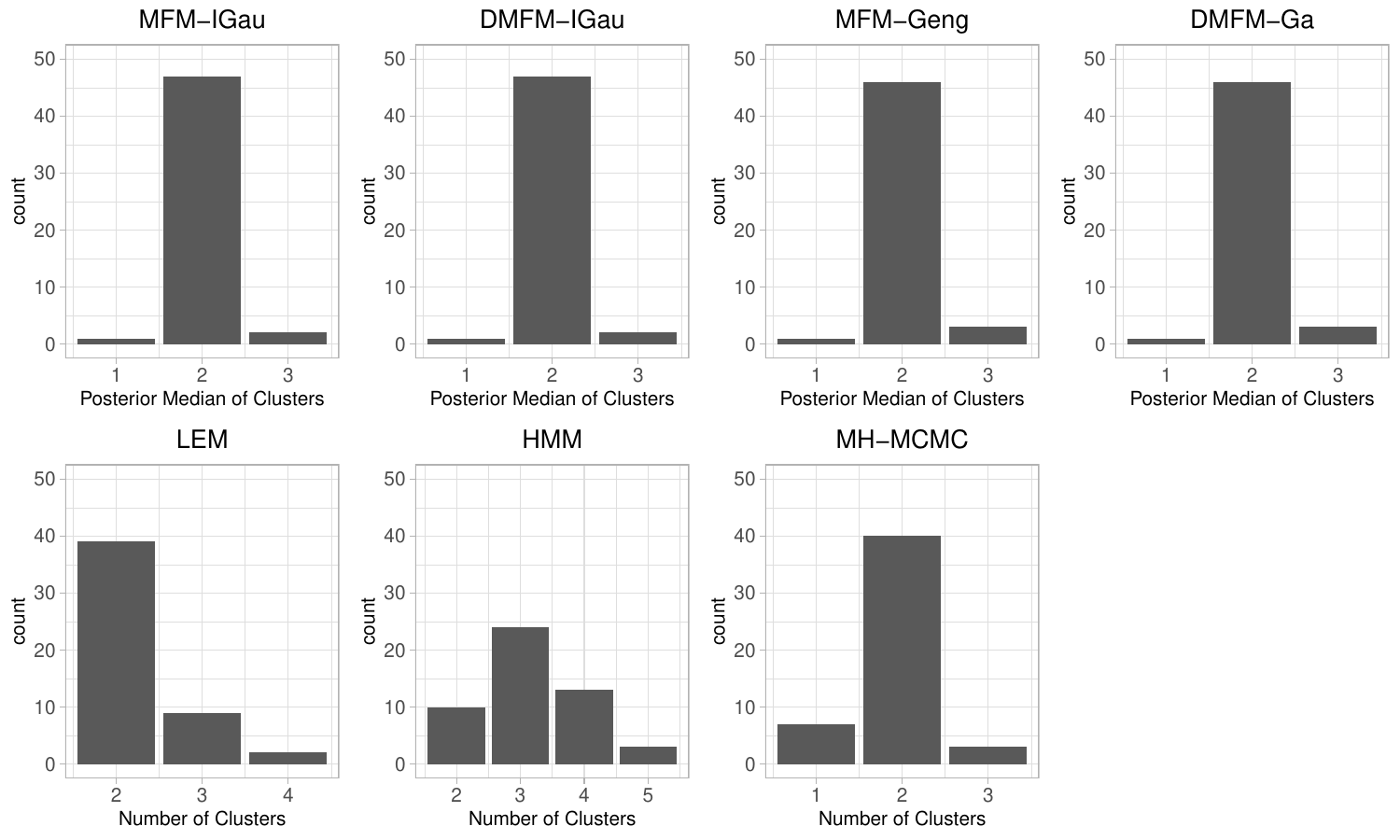}
    \caption{Histograms of estimated number of clusters in Section~4.3.1. From left to right: MFM-IGau, DMFM-IGau, MFM-Geng, DMFM-Ga, LEM, HMM and MH-MCMC.}
    \label{fig:bar_4_3_1_res2}
\end{figure}

As discussed in \cite{geng2019probabilistic}, MFM-Geng itself is a highly competitive community detection method and often outperforms several classical approaches under the experimental setting considered in Section~4.3.1. To further illustrate this point, we conducted additional numerical experiments by including the leading eigenvector method (LEM) \citep{PhysRevE.74.036104}, hierarchical modularity measure (HMM) \citep{Blondel_2008}, and MH-MCMC \citep{PhysRevLett.117.078301} as benchmark methods. LEM identifies dense community structures using the leading eigenvector of the modularity matrix derived from the adjacency matrix, thereby detecting relatively dense subgraphs. HMM is a heuristic method that optimizes modularity to detect dense community structures. MH-MCMC is a Bayesian method that updates $k$ via a Metropolis–Hastings step based on the so-called heat-bath algorithm \citep{newman1999monte}. The results for all MFM- and DMFM-based methods are identical to those reported in the main text. As shown in Figure~\ref{fig:bar_4_3_1_res2}, MFM-IGau, DMFM-IGau, MFM-Geng, and DMFM-Ga estimate the true number of clusters $k_{\text{true}} = 2$ more stably than LEM, HMM, and MH-MCMC. Furthermore, Table~\ref{tab:table_4_3_1_res2} shows that although LEM and HMM achieve high modularity values because they directly optimize modularity, their clustering accuracy is not necessarily superior to that of the Bayesian MFM-based approaches. In contrast, MH-MCMC consistently exhibits inferior performance compared with MFM-IGau, DMFM-IGau, MFM-Geng, and DMFM-Ga. Taken together, although MFM-IGau and DMFM-IGau do not universally outperform MFM-Geng and DMFM-Ga in terms of cluster number estimation, ARI, or modularity, they achieve comparable statistical performance while offering clear computational advantages. In addition, their ability to provide stable posterior support for the true number of communities further demonstrates the practical effectiveness of the proposed methods.

\begin{table}[htbp]
    \caption{Results of community detection for LEM, HMM and MH-MCMC over 50 simulated networks: ARI and Mod. Standard deviations are given in parentheses, where results of MH-MCMC are posterior means.}
    \begin{center}
    \begin{tabular}{lccccccc} 
    \hline
    & MFM-IGau & DMFM-IGau & MFM-Geng & DMFM-Ga & LEM & HMM & MH-MCMC \\ \hline
     ARI & $0.853$ & $0.855$ & $0.857$ & $0.859$ & $0.829$ & $0.659$ & $0.769$ \\
    & $(0.85)$ & $(0.084)$ & $(0.083)$ & $(0.083)$ & $(0.140)$ & $(0.201)$ & $(0.077)$ \\\hline
     Mod & $0.252$ & $0.258$ & $0.251$ & $0.253$ & $0.272$ & $0.251$ & $0.202$ \\
& $(0.060)$ & $(0.052)$ & $(0.057)$ & $(0.058)$ & $(0.022)$ & $(0.022)$ & $(0.086)$ \\\hline
\end{tabular}
\end{center}
\label{tab:table_4_3_1_res2}
\end{table}

Finally, we discuss the computational cost of updating the label variables \(c\). In our implementation, to ensure reproducibility of \cite{geng2019probabilistic} and fairness in comparison, we built our codes based on the algorithm in \cite{geng2019probabilistic} and the publicly available \textsf{R} code. The label update step in network models has a computational complexity on the order of $O(n^2)$, and this order does not change depending on whether a restaurant process mechanism is used. However, the constant factor (i.e., the per-iteration cost) differs substantially. As shown in Table~\ref{tab:time}, the label update step accounts for the vast majority of the total runtime, meaning that the form of the label update largely determines the overall computation time. Unlike methods based on a restaurant-process representation, our proposed algorithms do not require conditional branching during label updates to determine whether a new cluster is created or an existing cluster is removed. This enables more efficient label updates and is likely responsible for the substantial reduction in computation time.

\begin{table}[tbp]
    \caption{Comparison of the total MCMC run time and the total label-update time (one-shot) for MFM-IGau and MFM-Geng in the numerical experiment of Section 4.3.1.}
    \begin{center}
    \begin{tabular}{lcc} 
     \hline
    & MFM-IGau & MFM-Geng  \\ \hline
    total MCMC run time (sec) & $198.103$ & $638.680$\\ \hline
    total label update time (sec) & $153.519$ & $610.583$\\\hline
\end{tabular}
\end{center}
\label{tab:time}
\end{table}

\end{document}